\documentclass[%
 aip,
 amsmath,amssymb,
 reprint,%
]{revtex4-1}

\usepackage{graphicx}
\usepackage{dcolumn}
\usepackage{bm}
\usepackage{cleveref}

\usepackage[utf8]{inputenc}
\usepackage[T1]{fontenc}
\usepackage{threeparttable}
\usepackage[abbreviations]{glossaries-extra}
\usepackage{mathptmx}
\usepackage{etoolbox}
\usepackage{color}
\usepackage{CJKutf8}
\usepackage{xcolor}
\usepackage{soul}

\soulregister\cite7
\soulregister\ref7
\soulregister\eqref7
\soulregister\pageref7

\makeatletter
\def\@email#1#2{%
 \endgroup
 \patchcmd{\titleblock@produce}
  {\frontmatter@RRAPformat}
  {\frontmatter@RRAPformat{\produce@RRAP{*#1\href{mailto:#2}{#2}}}\frontmatter@RRAPformat}
  {}{}
}%
\makeatother
\begin{document}

\preprint{AIP/123-QED}

\title[]{Towards pore-scale simulation of combustion in porous media using a low-Mach hybrid lattice Boltzmann/finite difference solver}
\author{S.A.~Hosseini}%
\email{seyed.hosseini@ovgu.de.}
\affiliation{Laboratory of Fluid Dynamics and Technical Flows, University of Magdeburg ``Otto von Guericke'', D-39106 Magdeburg, Germany.}%
\affiliation{Department of Mechanical and Process Engineering, ETH Z\"urich, 8092 Z\"urich, Switzerland.}%
\author{D.~Th\'evenin}
\affiliation{Laboratory of Fluid Dynamics and Technical Flows, University of Magdeburg ``Otto von Guericke'', D-39106 Magdeburg, Germany.}%

\newabbreviation{cfd}{CFD}{computational fluid dynamics}
\newabbreviation{piv}{PIV}{particle image velocimetry}
\newabbreviation{chmrt}{CHMRT}{central Hermite multiple relaxation time}
\newabbreviation{srt}{SRT}{single relaxation time}
\newabbreviation{lbm}{LBM}{lattice Boltzmann method}

\date{\today}

\begin{abstract}
A hybrid numerical model previously developed for combustion simulations is extended in this article to describe flame propagation and stabilization in porous media. The model, with a special focus on flame/wall interaction processes, is validated via corresponding benchmarks involving flame propagation in channels with both adiabatic and constant-temperature walls. Simulations with different channel widths show that the model can correctly capture the changes in flame shape and propagation speed as well as the dead zone and quenching limit, as found in channels with cold walls. The model is further assessed considering a pseudo 2-D porous burner involving an array of cylindrical obstacles at constant temperature, investigated in a companion experimental study. Furthermore, the model is used to simulate pore-scale flame dynamics in a randomly-generated 3-D porous media. Results are promising, opening the door for future simulations of flame propagation in realistic porous media.
\end{abstract}

\maketitle
\section{\label{sec:introduction}Introduction}
Rapid depletion of fossil fuel resources and related pollutant emissions are a consequence of their widespread and abundant use in most areas of industry and technology~\cite{mujeebu_applications_2009,mujeebu_combustion_2009,mujeebu_trends_2010}. Motivated by these two issues, the search for more efficient and eco-friendly energy production technologies and their implementation at the industrial level is growing by the day. Combustion in porous media has been proven to be one promising route to tackle some of the previously-cited challenges. For burners, the concept of porous media can result in high power densities, increased power dynamic range, and low emissions of NO and ${\rm CO}_2$~\cite{trimis_combustion_1996}. This is, for the most part, the consequence of the presence of a solid porous matrix which has higher levels of heat capacity, conductivity, and emissivity as compared to the gaseous phase. The concept of combustion in porous media is also present in other eco-friendly technologies, for instance in packed bed reactors with Chemical Looping Combustion that allow for efficient separation of ${\rm CO}_2$~\cite{siriwardane_investigation_2016,shirzad_moving_2019}. Similar challenges involving intense flame/wall interactions are faced in meso- and micro-combustion found in corresponding burners developed within the context of micro electro-mechanical systems~\cite{shirsat_review_2011,maruta_micro_2011}. Given the pronounced impact of flame/solid interactions, the further development of such technologies requires a better understanding of flame/wall interaction dynamics. For this purpose, it is essential to develop numerical models that are able to properly capture such physics with a sufficient level of accuracy.\\
The topic of flame/wall interaction has been tackled in a variety of articles in the past decades, starting with investigations of head-on quenching~\cite{Poinsot}, mostly to quantify wall heat flux~\cite{Cuenot}. Such interesting investigations have been going on up to now, involving additional configurations and aspects as well as a variety of fuels~\cite{Hasse1,Hasse2}.
Even more relevant for the present investigations are flames propagating in narrow channels. Corresponding publications and results presented therein point to the very rich physics of the flame front when propagating in such a channel, see for instance~\cite{pizza_dynamics_2008,pizza_dynamics_2008-1,pizza_three-dimensional_2010,bioche_premixed_2018}. Depending on the ratio of the channel diameter to the flame thickness and on the type of thermal boundary condition at the wall the flame front can take on a wide variety of shapes, most notably, the so-called tulip shape~\cite{bioche_premixed_2018}. Extending further this line of research, flame propagation within porous media has also been studied with different levels of complexity, starting with academic configurations in \cite{sahraoui_direct_1994}. These preliminary studies led the authors to the conclusion that in the context of flame propagation in porous media, different flame propagation speeds exist, which is in agreement with the different propagation modes observed for flame propagation in channels. While volume-averaged approaches appear to be a cost-efficient tool for simulations of large-size, realistic systems, these observations clearly show the necessity of direct pore-scale simulations for a better understanding of the interaction process.\\
To the authors' knowledge, apart from~\cite{sawant_consistent_2022} where the authors model flame propagation in straight channels and ~\cite{lei_study_2021} where authors discuss specifically coal combustion, all studies targeting combustion applications in porous media and configurations dominated by flame/wall interactions have been carried out using classical, discrete solvers for the Navier-Stokes-Fourier equations, coupled to balance equations for the individual species. 
In the low-Mach number limit, to alleviate the limitation in time-step resulting from the presence of acoustic modes, most such solvers rely on the so-called zero-Mach approximation~\cite{majda_derivation_1985}, which by virtue of the Helmholtz decomposition of the velocity field brings the Poisson equation into the scheme, see for instance~\cite{abdelsamie_towards_2016}. The elliptic Poisson equation is well-known to be the computational bottleneck of incompressible Navier-Stokes models. To solve this issue, different approaches such as Chorin's artificial compressibility method (ACM)~\cite{chorin_numerical_1997} replacing the Poisson equation with a hyperbolic equation for the pressure have been proposed for incompressible flows.\\
The lattice Boltzmann method (LBM), which emerged in the literature in the late 80's~\cite{succi_lattice_2002}, has now achieved widespread success. This is in particular due to the fully hyperbolic nature of all involved equations. In addition, and as an advantage over ACM, normal acoustic modes are also subject to dissipation and, therefore, are governed by a parabolic partial differential equation allowing the LBM to efficiently tackle unsteady flows. Following up on the same idea, we recently proposed an algorithm for low-Mach thermo-compressible flows based on the lattice Boltzmann method~\cite{hosseini_hybrid_2019,hosseini_development_2020,hosseini_low-mach_2020}. Different from other LBM approaches proposed in recent years for combustion simulation~\cite{feng_lattice-boltzmann_2018,lei_study_2021,sawant_consistent_2022}, this scheme is specifically tailored for the low-Mach regime. While this model has been successfully used for large-eddy simulations (LES) of flames in complex geometries, in particular swirl burners~\cite{hosseini_low_2022}, detailed interactions between flame fronts and walls have not been considered in detail up to now, since they did not play a central role for the considered systems.\\
In this study a corresponding validation of the solver is proposed, including boundary conditions for curved walls. Configurations of increasing complexity are considered, such as flame propagation in narrow channels of different widths involving different thermal boundary conditions, as well as combustion in a reference 2-D packed bed reactor corresponding to a companion experimental study. Note that the so-called pores considered in the present study are large, being indeed inter-particle spaces at the millimeter or centimeter scale, and not restricted to a few micrometers, as found in many other applications. In this article, the terms pore and inter-particle space are used interchangeably to designate the same configuration.\\
After a brief refresher of the model itself, along with its multiple relaxation time (MRT) cumulants realization, a discussion of the boundary conditions is proposed for both the lattice Boltzmann and the finite-difference (FD) solvers. Afterwards, results from the different validation cases are presented and discussed, before conclusion.
\section{Theoretical background\label{sec:numericalMethod}}
\subsection{\label{subsec:governing_equations} Governing equations}
The model used here and detailed in the next subsections targets the low-Mach approximation to describe thermo-compressible reacting flows~\cite{poinsot_theoretical_2005}. The species mass balance equation reads in non-conservative form:
\begin{equation}\label{eq:species_balance}
 \partial_t Y_k + \bm{u}\cdot\bm{\nabla}Y_k + \frac{1}{\rho}\bm{\nabla}\cdot\rho\bm{V}_k Y_k = \frac{\dot{\omega}_k}{\rho},
\end{equation}
where $Y_k$ is the $k^{\rm th}$ species mass fraction, $\rho$ the local density, $\bm{u}$ the mixture velocity, and $\dot{\omega}_k$ the source term due to chemical reactions. The mass flux due to diffusion, $Y_k\bm{V}_k$, is given by:
\begin{equation}\label{eq:species_diffusion_speed}
 Y_k\bm{V}_k = -\frac{D_k W_k}{W}\bm{\nabla}X_k + Y_k\sum_{k'=1}^{N_{\rm sp}} \frac{D_{k'} W_{k'}}{W}\bm{\nabla}X_{k'}
\end{equation}
where $X_k$, $W_k$ and $D_k$ are respectively the $k^{\rm th}$ species mole fraction, molar mass and mixture-averaged diffusion coefficient. $W$ is the mixture molar mass. The second term corresponds to the correction velocity ensuring local conservation of total mass (i.e., $\sum_{k=1}^{N_{\rm sp}} Y_k\bm{V}_k=0$).

The momentum balance equation (Navier-Stokes) reads:
\begin{equation}\label{eq:momentum_balance}
 \partial_t (\rho \bm{u}) + \bm{\nabla}\cdot(\rho\bm{u}\otimes\bm{u}) + \bm{\nabla}\cdot\bm{S} = 0,
\end{equation}
where the stress is:
\begin{equation}\label{eq:stress_tensor}
 \bm{S} = P_h\bm{I} - \mu\left(\bm{\nabla}\bm{u} + \bm{\nabla}\bm{u}^{t} - \frac{2}{D}\bm{\nabla}\cdot\bm{u}\bm{I}\right) - \eta \bm{\nabla}\cdot\bm{u}\bm{I}.
\end{equation}
in which $\mu$ and $\eta$ are the mixture-averaged dynamic and bulk viscosity coefficients and $P_h$ is the hydrodynamic pressure tied to the total pressure as $P=P_0+ P_h$, with $P_0$ the uniform thermodynamic pressure. 
The employed closure for the hydrodynamic pressure $P_h$ reads:
\begin{equation}\label{eq:pressure_equation}
 \frac{1}{\rho c_s^2}\partial_t P_h + \bm{\nabla}\cdot\bm{u} = \Lambda,
\end{equation}
where $c_s$ is the characteristic propagation speed of normal modes, also known as sound speed. At the difference of a truly compressible model, here $c_s$ is not necessarily the physical sound speed. Using the continuity equation and the ideal gas mixture equation of state, one gets:
\begin{equation}
 \Lambda = \frac{\partial_t T + \bm{u}\cdot\bm{\nabla}T}{T} \\ + \sum_{k=1}^{N_{\rm sp}}\frac{W}{W_k}\left(\partial_t Y_k + \bm{u}\cdot\bm{\nabla}Y_k\right).
\end{equation}
Finally the energy balance equation is given by
\begin{multline}\label{eq:energy_balance}
 \rho c_p\left(\partial_t T + \bm{u}\cdot\bm{\nabla}T\right) - \bm{\nabla}\cdot(\lambda \bm{\nabla}T) \\ + \rho\left(\sum_{k=1}^{N_{\rm sp}} c_{p_k} Y_k \bm{V}_k\right)\cdot\bm{\nabla}T = \dot{\omega}_T,
\end{multline}
where $c_{p_k}$ and $c_p$ are respectively the $k^{\rm th}$ species and the mixture specific heat capacities and $\lambda$ is the thermal diffusion coefficient.\\
One point that is to be noted is the difference of the current low-Mach set of equations with the zero-Mach model of Majda and the low-Mach model of Toutant; Setting $c_s$ to be the real sound speed in Eq.~\ref{eq:pressure_equation} reduces it to that of \cite{toutant_general_2017}, but now for a multi-species reacting system. On the other hand, in the limit of $c_s\rightarrow\infty$ one ends up with Majda's zero-Mach limit~\cite{majda_derivation_1985}, i.e. $\bm{\nabla}\cdot\bm{u} = \Lambda$. A detailed perturbation analysis of this system would be interesting but will be left for future publications. In the next section the lattice Boltzmann model used to recover the corresponding hydrodynamic limit is briefly introduced.
\subsection{Lattice Boltzmann model\label{subsec:lattice_boltzmann}}
To solve the low-Mach aerodynamic equations, we use a lattice Boltzmann model that we have developed in previous works~\cite{hosseini_hybrid_2019,hosseini_low-mach_2020,hosseini_development_2020}:
\begin{equation}\label{eq:LBM_equation}
 g_i(\bm{r}+\bm{c}_i\delta r,t+\delta t) - g_i(\bm{r},t) = \Omega_i + \delta t\Xi_i,
\end{equation}
where $g_i$ are discrete populations, $\bm{c}_i$ corresponding discrete velocities, $\bm{r}$ and $t$ the position in space and time, $\delta t$ the time-step size and
\begin{equation}
 \Xi_i = c_s^2\left(f^{\rm eq}_i/\rho -w_i\right)\left(\bm{c}_i-\bm{u}\right)\cdot\bm{\nabla}\rho
 + w_i\rho c_s^2 \Lambda.
\end{equation}
Here, $W_k$ is the molar mass of species $k$ and $W$ the average molar mass, $N_{\rm sp}$ the number of species, $w_i$ the weights associated to each discrete velocity in the lattice Boltzmann solver and $c_s$ the lattice sound speed tied to the time-step and grid size $\delta r$ as $c_s=\delta r/\sqrt{3}\delta t$. The equilibrium distribution function, $f^{\rm eq}_i$, is given by:
\begin{equation}
 f^{\rm eq}_i = w_i\rho\left(1+\frac{\bm{c}_i.\bm{u}}{c_s^2} + \frac{{(\bm{c}_i.\bm{u})}^2}{2c_s^4} - \frac{\bm{u}^2}{2c_s^2}\right).
\end{equation}
The collision term $\Omega_i$ is defined as:
\begin{equation}
 \Omega_i = -\omega_s\left(g_i-g_i^{\rm eq}\right),
\end{equation}
where 
\begin{equation}
 g_i^{\rm eq} = w_i(P_h - \rho c_s^2) + c_s^2 f_i^{\rm eq},
\end{equation}
and $P_h$ is the hydrodynamic pressure. In the present study first-neighbour stencils based on third-order quadratures are used, i.e. D2Q9 and D3Q27. The hydrodynamic pressure and momentum are computed as moments of the distribution function $g_i$:
\begin{subequations}
\begin{align}
	P_h &= \sum_{i=1}^{Q} g_i + \frac{\delta t}{2}\rho c_s^2\Lambda,\\
	\rho\bm{u} &= \frac{1}{c_s^2}\sum_{i=1}^{Q} \bm{c}_i g_i.
	\end{align}
\label{eq:moments_PDF}
\end{subequations}
This lattice Boltzmann model recovers the previously introduced pressure evolution equation along with the Navier-Stokes equation. In the viscous stress tensor deviations from Galilean invariance are limited to third order.
\subsection{Implementation of the Multiple Relaxation Times (MRT) collision operator\label{subsec:mrt}}
In the context of the present study, following our proposals for both multi-phase and multi-species flows~\cite{hosseini_lattice_2021,hosseini_low_2022}, the Cumulants-based operator is used~\cite{geier_cumulant_2015}. The post-collision populations $g_i^{*}$ are computed as:
\begin{equation}
    g_i^{*} = \rho c_s^2 {f^{'}_i}^{*} + \frac{\delta t}{2}\Xi_i,
\end{equation}
where the post-collision pre-conditioned populations ${f^{'}_i}^{*}$ are:
\begin{equation}
     {f^{'}_i}^{*} = \mathcal{M}^{-1}\left(\mathcal{I} - \mathcal{W}\right)\mathcal{K}^{'} + \mathcal{M}^{-1}\mathcal{W}\mathcal{K}^{'},
\end{equation}
In this equation, $\mathcal{M}$ is the moments transform matrix from pre-conditioned populations to the target momentum space, $\mathcal{I}$ the identity matrix and $\mathcal{W}$ the diagonal relaxation frequencies matrix
\begin{equation}
    \mathcal{W}={\rm diag}(\omega_0, \omega_x, \omega_y, ..., \omega_{xxyyzz}),
\end{equation}
where the operator ${\rm diag}$ is defined as:
\begin{equation}
    {\rm diag}(\bm{A}) = (\bm{A}\otimes\bm{1})\circ \mathcal{I},
\end{equation}
with $\bm{A}$ a given vector and $\bm{1}$ a vector with elements 1. The relaxation frequencies of second-order shear moments, e.g. $xy$ (here shown with $\omega_s$ for the sake of readability) are defined as:
\begin{equation}
    \omega_s = \frac{\nu}{c_s^2\delta t} + \frac{1}{2},
\end{equation}
where $\nu$ is the local effective kinematic viscosity. Prior to transformation to momentum space the populations are pre-conditioned as:
\begin{equation}
    f^{'}_i = \frac{1}{\rho c_s^2} g_i + \frac{\delta t}{2\rho c_s^2}\Xi_i.
\end{equation}
This pre-conditioning accomplishes two tasks, 1) normalizing the populations with the density -- and thus eliminating the density-dependence of the moments --, and 2) introducing the first half of the source term. As such the moments $\mathcal{K}^{'}$ are computed as:
\begin{equation}
    \mathcal{K}^{'}_{j} = \mathcal{M}_{ij} f^{'}_i.
\end{equation}
The  Cumulants $\mathcal{K}_j$ are computed from the central moments of the distribution function, these central moments being defined as:
\begin{equation}
    \widetilde{\Pi}^{'}_{x^p y^q z^r} = \sum_i {\left(c_{i,x} - u_x\right)}^p {\left(c_{i,y} - u_y\right)}^q {\left(c_{i,z} - u_z\right)}^r f^{'}_\alpha.
\end{equation}
As noted in \cite{geier_cumulant_2015}, up to order three Cumulants are identical to their central moments counter-parts. At higher orders they are computed as:
	\begin{subequations}
		\begin{align}
		\mathcal{K}^{'}_{xxyz} &= \widetilde{\Pi}^{'}_{xxyz} - \widetilde{\Pi}^{'}_{xx}\widetilde{\Pi}^{'}_{yz} - 2\widetilde{\Pi}^{'}_{xy}\widetilde{\Pi}^{'}_{xz},\\
		\mathcal{K}^{'}_{xxyy} &= \widetilde{\Pi}^{'}_{xxyy} - \widetilde{\Pi}^{'}_{zz}\widetilde{\Pi}^{'}_{xyy} - 2\widetilde{\Pi}^{'}_{xy}\widetilde{\Pi}^{'}_{xy},\\
		\mathcal{K}^{'}_{xyyzz} &= \widetilde{\Pi}^{'}_{xyyzz} - \widetilde{\Pi}^{'}_{yyzz}\widetilde{\Pi}^{'}_{xyy} - \widetilde{\Pi}^{'}_{yy} \widetilde{\Pi}^{'}_{xzz} \nonumber\\ &- 4\widetilde{\Pi}^{'}_{yz} \widetilde{\Pi}^{'}_{xyz} - 2\widetilde{\Pi}^{'}_{xz}\widetilde{\Pi}^{'}_{yyz} - 2\widetilde{\Pi}^{'}_{xy}\widetilde{\Pi}^{'}_{yzz},\\
		\mathcal{K}^{'}_{xxyyzz} &= \widetilde{\Pi}^{'}_{xxyyzz} - 4\widetilde{\Pi}^{'}_{xyz}\widetilde{\Pi}^{'}_{xyz} -
		\widetilde{\Pi}^{'}_{xx} \widetilde{\Pi}^{'}_{yyzz} \nonumber\\ &- \widetilde{\Pi}^{'}_{yy} \widetilde{\Pi}^{'}_{xxzz} - \widetilde{\Pi}^{'}_{zz}\widetilde{\Pi}^{'}_{yyzz} - 4\widetilde{\Pi}^{'}_{yz}\widetilde{\Pi}^{'}_{xxyz}\nonumber\\
		&- 4\widetilde{\Pi}^{'}_{xz}\widetilde{\Pi}^{'}_{xyyz} - 4\widetilde{\Pi}^{'}_{xy}\widetilde{\Pi}^{'}_{xyzz} - 2\widetilde{\Pi}^{'}_{xyy}\widetilde{\Pi}^{'}_{xzz}\nonumber\\
		&- 2\widetilde{\Pi}^{'}_{xxy}\widetilde{\Pi}^{'}_{yzz} - 2\widetilde{\Pi}^{'}_{xxz}\widetilde{\Pi}^{'}_{yyz} + 16\widetilde{\Pi}^{'}_{xy}\widetilde{\Pi}^{'}_{xz}\widetilde{\Pi}^{'}_{yz}\nonumber\\
		&+ 4 \widetilde{\Pi}^{'}_{xz}\widetilde{\Pi}^{'}_{xz}\widetilde{\Pi}^{'}_{yy} + 4\widetilde{\Pi}^{'}_{yz}\widetilde{\Pi}^{'}_{yz}\widetilde{\Pi}^{'}_{xx} + 4\widetilde{\Pi}^{'}_{xy}\widetilde{\Pi}^{'}_{xy}\widetilde{\Pi}^{'}_{zz}\nonumber\\ &+ 2\widetilde{\Pi}^{'}_{xx}\widetilde{\Pi}^{'}_{yy}\widetilde{\Pi}^{'}_{zz}.
		\end{align}
		\label{Eq:cumulants}
	\end{subequations}
Given that the Cumulants of the equilibrium distribution functions are equal to zero, the post-collision Cumulants are readily obtained as:
\begin{equation}
    \mathcal{K}^{'*}_{j} = \left(1-\omega_j\right) \mathcal{K}^{'}_{j},
\end{equation}
with $\omega_j$ the relaxation frequency of Cumulant $j$.
After collision, the Cumulants $ \mathcal{K}^{'*}_{j}$ have to be transformed back into populations $f_i^{'*}$. The first step, as for the forward transformation is to get the corresponding central moments. Given that up to order three central moments and Cumulants are the same, we only give here the backward transformation of higher-order moments:
	\begin{subequations}
		\begin{align}
		\widetilde{\Pi}^{'*}_{xxyz} &= \mathcal{K}^{'*}_{xxyz} + \widetilde{\Pi}^{'*}_{xx}\widetilde{\Pi}^{'*}_{yz} + 2\widetilde{\Pi}^{'*}_{xy}\widetilde{\Pi}^{'*}_{xz}\\
        \widetilde{\Pi}^{'*}_{xxyy} &= \mathcal{K}^{'*}_{xxyy} + \widetilde{\Pi}^{'*}_{xx}\widetilde{\Pi}^{'*}_{yy} + 2\widetilde{\Pi}^{'*}_{xy}\widetilde{\Pi}^{'*}_{xy}\\
        \widetilde{\Pi}^{'*}_{xyyzz} &=\mathcal{K}^{'*}_{xyyz} + \widetilde{\Pi}^{'*}_{zz}\widetilde{\Pi}^{'*}_{xyy} + \widetilde{\Pi}^{'*}_{yy}\widetilde{\Pi}^{'*}_{xzz} + 4\widetilde{\Pi}^{'*}_{yz}\widetilde{\Pi}^{'*}_{xyz}\nonumber\\
         &+ 2\widetilde{\Pi}^{'*}_{xz}\widetilde{\Pi}^{'*}_{yyz} + 2\widetilde{\Pi}^{'*}_{xy}\widetilde{\Pi}^{'*}_{yzz}\\
         \widetilde{\Pi}^{'*}_{xxyyzz} &= \mathcal{K}^{'*}_{xxyyzz} + 4\widetilde{\Pi}^{'*}_{xyz}\widetilde{\Pi}^{'*}_{xyz} + \widetilde{\Pi}^{'*}_{xx}\widetilde{\Pi}^{'*}_{yyzz}\nonumber\\
          &+ \widetilde{\Pi}^{'*}_{yy}\widetilde{\Pi}^{'*}_{xxzz} + \widetilde{\Pi}^{'*}_{zz}\widetilde{\Pi}^{'*}_{xxyy} + 4\widetilde{\Pi}^{'*}_{yz}\widetilde{\Pi}^{'*}_{xxyz}\nonumber\\
           &+ 4\widetilde{\Pi}^{'*}_{xz}\widetilde{\Pi}^{'*}_{xyyz} + 4\widetilde{\Pi}^{'*}_{xy}\widetilde{\Pi}^{'*}_{xyzz} + 2\widetilde{\Pi}^{'*}_{xyy}\widetilde{\Pi}^{'*}_{xzz}\nonumber\\
            &+ 2\widetilde{\Pi}^{'*}_{xxy}\widetilde{\Pi}^{'*}_{yzz} + 2\widetilde{\Pi}^{'*}_{xxz}\widetilde{\Pi}^{'*}_{yyz} + 16\widetilde{\Pi}^{'*}_{xy}\widetilde{\Pi}^{'*}_{xz}\widetilde{\Pi}^{'*}_{yz}\nonumber\\
             &+ 4\widetilde{\Pi}^{'*}_{xz}\widetilde{\Pi}^{'*}_{xyyz}+4\widetilde{\Pi}^{'*}_{xy}\widetilde{\Pi}^{'*}_{xyzz}+4\widetilde{\Pi}^{'*}_{yz}\widetilde{\Pi}^{'*}_{xxyz}\nonumber\\
              &+ 2\widetilde{\Pi}^{'*}_{xx}\widetilde{\Pi}^{'*}_{yy}\widetilde{\Pi}^{'*}_{zz}.
		\end{align}
		\label{Eq:Backcumulants}
	\end{subequations}
Once central moments have been obtained the inverse of the central moments transform tensor is used to compute the corresponding populations.
\subsection{Solver for species and energy balance equations\label{subsec:energy_species_discrete}}
In the context of the present study the species and energy balance laws (Eqs.~\ref{eq:species_balance} and \ref{eq:energy_balance}) are solved using finite differences. To prevent the formation of Gibbs oscillations at sharp interfaces, convective terms are discretized using a third-order weighted essentially non-oscillatory (WENO) scheme while diffusion terms are treated via a fourth-order central scheme. Near boundary nodes, to prevent any nonphysical interaction of the smoothness indicator with ghost nodes, a centered second-order scheme is used to discretize the convection term. Global mass conservation of the species balance equation, i.e. $\sum_k Y_k = 1$, while naturally satisfied for classical discretizations of the convection term, for instance in 1-D:
\begin{equation}
    \frac{u_x}{2\delta r}\sum_k \left[Y_k(x+\delta r) - Y_k(x-\delta r)  \right] = 0.
\end{equation}
is not necessarily satisfied for WENO schemes, as coefficients weighing contributions of each stencil are not the same for all species. To guarantee conservation of overall mass the concept of correction speed is used as for the diffusion model; Representing the discretization via an operator $\mathcal{L}$ the discrete convection term is computed as:
\begin{equation}
    \bm{u}\cdot\bm{\nabla}Y_k = \bm{u}\cdot\left[\mathcal{L}(\bm{\nabla}Y_k) - Y_k\sum_{k'}\mathcal{L}(\bm{\nabla}Y_{k'})\right]
\end{equation}
which -- once summed up over all species -- gives:
\begin{equation}
    \sum_k \bm{u}\cdot\bm{\nabla}Y_k = \bm{u}\cdot\left[\sum_k\mathcal{L}(\bm{\nabla}Y_k) - \sum_{k'}\mathcal{L}(\bm{\nabla}Y_{k'})\right] = 0. 
\end{equation}
All equations are discretized in time using a first-order Euler approach. Transport and thermodynamic properties of the mixture along with the kinetic scheme are taken into account via the open-source library Cantera, coupled to our in-house solver ALBORZ~\cite{hosseini_development_2020}. Details of the coupling can be found in \cite{hosseini_weakly_2020}.
\section{\label{sec:BC}Boundary conditions}
\subsection{Lattice Boltzmann solver\label{subsec:LB_BC}}
In the context of the present study three types of boundary conditions are needed for the lattice Boltzmann solver, namely wall, inflow, and outflow boundary conditions. A brief overview of these boundary conditions is given in what follows.\\
\par Solid boundaries are modeled using the half-way bounce-back scheme. For this purpose,  missing populations are computed as \cite{kruger_lattice_2017}:
\begin{equation}
    f_i\left(\bm{r},t+\delta t\right) = f^{*}_{\bar{i}}\left(\bm{r},t\right),
\end{equation}
where $f^{*}_{\bar{i}}$ is the post-collision population (prior to streaming) and $\bar{i}$ is the index of the particle velocity opposite that of $i$. To take into account wall curvature the interpolated half-way bounce back approach is used~\cite{bouzidi_momentum_2001}.
At a given boundary node $\bm{r}_f$, the missing incoming populations are computed as:
\begin{subequations}
	\begin{align}
		f_i(\bm{r}_f, t+\delta t) &= 2qf_{\bar{i}}(\bm{r}_f+\bm{c}_{\bar{i}}, t+\delta t)\nonumber \\  &+\left(1-2q\right)f_{\bar{i}}(\bm{r}_f, t+\delta t), \forall q<\frac{1}{2},\\
		f_i(\bm{r}_f, t+\delta t) &= \frac{1}{2q}f_{\bar{i}}(\bm{r}_f+\bm{c}_{\bar{i}}, t+\delta t)\nonumber \\  &+\frac{2q-1}{2q}f_{\bar{i}}(\bm{r}_f, t+\delta t), \forall q\geq\frac{1}{2},
		\end{align}
	\label{Eq:CE_moments_eq}
\end{subequations}
where $\bar{i}$ designates the direction opposite $i$ and $q$ reads:
\begin{equation}
    q = \frac{\lvert\lvert \bm{r}_f - \bm{r}_s\lvert\lvert}{\lvert\lvert\bm{c}_i \lvert\lvert},
\end{equation}
with $\bm{r}_s$ denoting the wall position along direction $i$.\\
\par For inlet boundary conditions a modified version of the half-way bounce-back scheme is used to impose a target inlet velocity vector $\bm{u}_{\rm in}$. To that end the missing populations are computed as:
\begin{equation}
    f_i\left(\bm{r},t+\delta t\right) = f^{*}_{\bar{i}}\left(\bm{r},t\right) + (w_i + w_{\bar{i}}) \rho_{\rm in} \bm{u}_{\rm in}\cdot \bm{c}_i.
\end{equation}
\par In addition to velocity boundary conditions, a modified non-reflecting version of the zero-gradient boundary condition is also employed~\cite{kruger_lattice_2017} at the outlet, as first introduced in~\cite{geier_cumulant_2015}. The missing populations at the outflow boundary are defined as:
\begin{multline}
    f_i\left(\bm{r},t+\delta t\right) = f_i\left(\bm{r}-\bm{n}\delta r,t\right) \left(c_s - \bm{u}(\bm{r},t)\cdot\bm{n}\right) \\ + f_i\left(\bm{r}\delta r,t\right) \left(\frac{\delta r}{\delta t} - c_s + \bm{u}(\bm{r},t)\cdot\bm{n}\right),
\end{multline}
where $\bm{n}$ is the outward-pointing unit vector normal to the boundary surface.
\subsection{Energy and Species fields}
\par In addition to the application of boundary conditions to the discrete populations, given that the model involves derivatives of macroscopic properties such as density, appropriate measures have to be taken.
\begin{figure}[!h]
	\centering
		 \includegraphics[width=0.6\columnwidth]{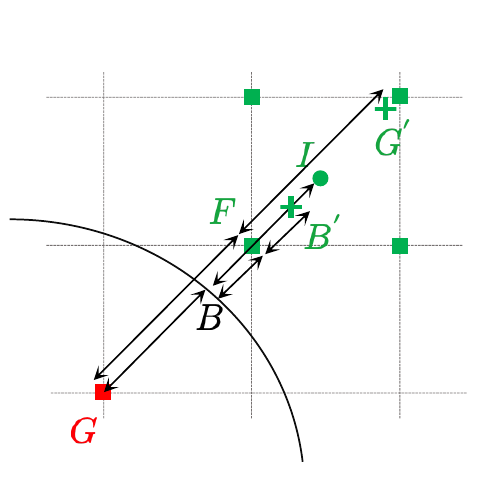}
\caption{Illustration of the ghost/image node approach for boundary conditions. The red point $G$ is outside the flow domain while green points are inside.}
\label{Fig:FD_ghost_BC}
\end{figure}
For the finite-difference solver and all terms involving this approximation, the boundary conditions are implemented via the image/ghost node method~\cite{pan_computation_2009,pan_simple_2010,baeza_high_2016}. Representing the macroscopic parameter of interest with the generic variable $\phi$, for a Dirichlet boundary condition for instance, one would have:
\begin{equation}
 \phi(B) = \phi_B,
\end{equation}
where $B$ refers to the position of the boundary, shown in Fig.~\ref{Fig:FD_ghost_BC}.
The \emph{virtual} field value $\phi(G)$ in the ghost node, the discrete grid-point outside the fluid domain neighboring the boundary (Fig.~\ref{Fig:FD_ghost_BC}) is computed as:
\begin{equation}
 \phi(G) = 2\phi_B - \phi(I),
\end{equation}
where $I$ is the image point in the fluid domain placed such that $\overline{GB}=\overline{BI}$ with both line segments perpendicular to the boundary interface. Since the image node does not necessarily fall on a grid-point it is reconstructed using data from neighboring grid points. For the reconstruction process to be robust with respect to the wall geometry, Shepard's inverse distance weighting is used~\cite{shepard_two-dimensional_1968}:
\begin{equation}
 \phi(r_I) = \sum_{j=1}^{N}w_j\phi(r_j),
\end{equation}
with:
\begin{equation}
w_j = \frac{d(r_I,r_j)^{-p}}{\sum_{j'=1}^{N}d(x_I,r_{j'})^{-p}},
\end{equation}
where $d(r_I,r_j)$ is the distance between points $I$ and $j$ and $p$ is a free parameter typically set to $p=2$. Note that:
\begin{equation}
    \sum_j w_j = 1.
\end{equation}
In order to obtain good precision, the field reconstruction at image points considers all fluid nodes neighboring $I$ such that:
\begin{equation}
    d(r_I,r_j) \leq 4\delta r,
\end{equation}
which comes at the additional cost of a wider data exchange layer between cores during parallelization.
\par Note that terms involving second-order derivatives such as the diffusion term in the energy and species balance equations also require an interpolation/reconstruction process on the diffusion coefficient. To avoid non-physical values, instead of using the previously computed properties, the coefficients at the ghost nodes are computed by applying the interpolation/reconstruction procedure directly to the transport properties.
\section{Validations and results\label{sec:results}}
\subsection{Premixed laminar flame acceleration in 2-D channels}
\begin{figure}[!h]
	\centering
		 \includegraphics[width=0.9\columnwidth]{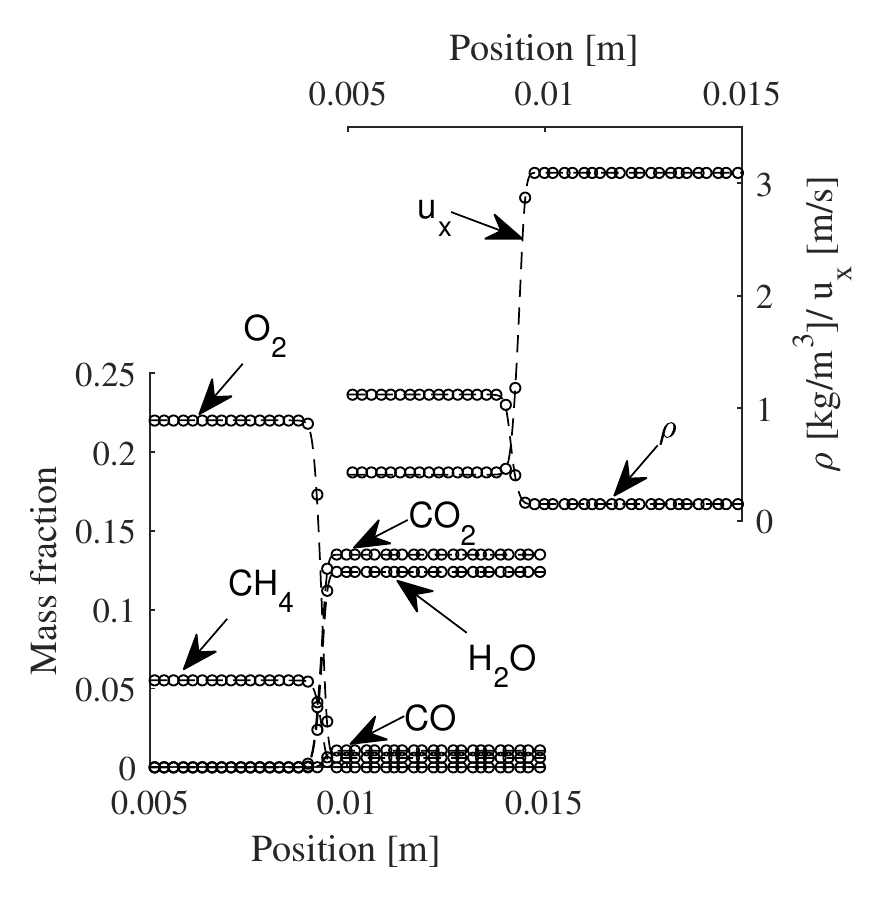}
\caption{Validation of stoichiometric methane/air flame against reference solver: The dashed lines are from Cantera, while the markers have been computed using ALBORZ.}
\label{Fig:1d_flame_data}
\end{figure}

\begin{figure}[h!]
	\centering
	\includegraphics[width=6cm,keepaspectratio]{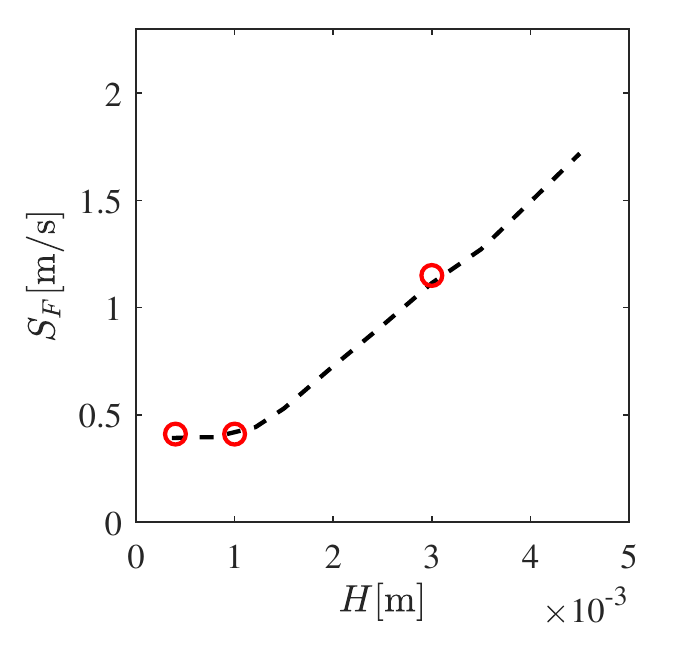}
	\caption{Comparison of flame propagation speed obtained for adiabatic walls from simulations with ALBORZ to results from~\cite{kim_numerical_2006} for different channel widths. Red circular markers are ALBORZ results while the black dashed line is data from ~\cite{kim_numerical_2006}.}
	\label{Fig:adiabatic_flame_speed}
\end{figure}

\begin{figure}[h!]
	\centering\includegraphics[width=9cm,keepaspectratio]{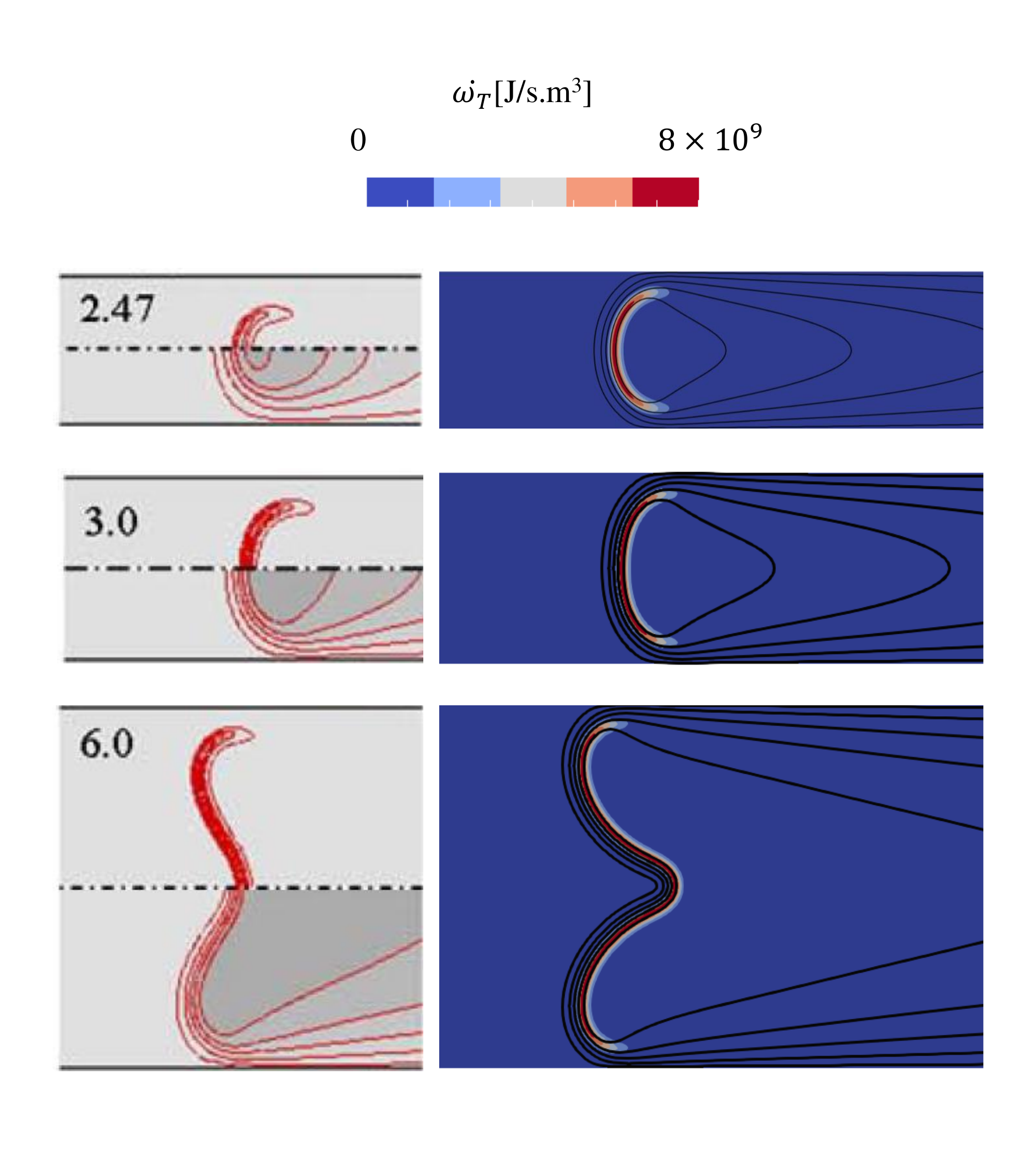}
	\caption{Comparison of flame shape obtained for isothermal walls obtained from simulations with ALBORZ (right half of the figure) to results from~\cite{kim_numerical_2006} (left half of the figure), with increasing channel height from top to bottom. The colors show the heat release rate on the right side, while the iso-contours in black in the right part represent the following isotherms: $\theta=\frac{T-T_{\rm in}}{T_{\rm ad}-T_{\rm in}}\in\{0.1,\, 0.3,\, 0.5,\, 0.7,\, 0.9\}$. Reference images showing iso-contours on the left (top half: heat release; bottom half: temperature) are reproduced from~\cite{kim_numerical_2006}.  Channel widths are set to be true to scale.}
	\label{Fig:isothermal_flame_shape}
\end{figure}

The proper interaction of flames with different wall boundary conditions (isothermal, adiabatic, known heat flux) while enforcing the no-slip condition for the flow is probably the most important step when extending a combustion solver to porous media applications. To that end, the propagation of premixed flames in narrow 2-D channels is first considered to verify that the proposed solver correctly captures the different flame front regimes.\\
Two configurations are considered: (a) Adiabatic and (b) constant-temperature channel walls. Given that the width of the channel, here written $H$, plays an important role to control flame front shape, heat exchange, as well as propagation speed, different cases with different channel widths have been computed. All configurations involve 2-D channels of height $H$ and length $L=20 H$. At the inflow (left end of the domain) a stoichiometric mixture of methane/air at temperature $T_{\rm in} = 300~{\rm K}$ is injected. The flow rate is dynamically set throughout all simulations to match the flame propagation speed, so as to ensure a globally static flame front within the numerical domain. The top and bottom boundaries are set to no-slip walls with either constant temperature, i.e. $T_w=T_{\rm in}$, or adiabatic boundary conditions for the temperature field. At the outlet a constant-pressure boundary condition is used. Note that for the inlet a 2-D Poiseuille distribution satisfying the target mass flow rate is implemented. To initialize all simulations profiles from the steady solution of a 1-D methane/air flame with the flame placed half-way in the domain are used and supplemented with the velocity distribution at the inlet.\\
 \paragraph{1-D free flame properties}
As a first step pseudo 1-D free flame simulations were run both using ALBORZ (coupled to Cantera) or Cantera (as standalone tool) using the BFER-2 two-step kinetic mechanism~\cite{franzelli_impact_2013}. The results obtained with both codes have been compared, as illustrated in Fig.~\ref{Fig:1d_flame_data}; the agreement is perfect for all species and all quantities. For this case, experimental measurements led to a flame propagation speed of $S_F=0.404~{\rm m}/{\rm s}$~\cite{elia2001laminar}, in excellent agreement with both solvers; ALBORZ predicts a laminar flame speed of 0.408 m/s.
Furthermore, to have a clear indication regarding resolution requirements, the thermal thickness $\delta_T$ defined as:
\begin{equation}
    \delta_T = \frac{T_{\rm ad} - T_{\rm in}}{dT/dx},
\end{equation}
where $T_{\rm ad}$ is the adiabatic flame temperature, was also computed. Simulations with ALBORZ led to $\delta_T = 328~\mu{\rm m}$, which is in very good agreement with the value reported in~\cite{kim_numerical_2006}. This indicates that for fully resolved simulations one should implement $\delta r<35~\mu{\rm m}$, in order to get 10 grid points within the flame front. For all channel simulations conducted in the present section $\delta r=20~\mu{\rm m}$ has been set. While larger grid-sizes would be sufficient for resolved simulations, as will be seen in next section, here we use a smaller grid-size to properly resolve the width of the smaller channel. Considering additionally the characteristic speed in the system the time-step size was then fixed to $\delta t=7.5\times10^{-8}~{\rm s}$, also satisfying all stability conditions regarding Fourier and CFL numbers for the hybrid solver.\\
\paragraph{Adiabatic walls}
For the first set of simulations the walls are set to be adiabatic. Three different channel widths are considered, i.e. $H\in\{0.4,\, 1,\, 3\} {\rm mm}$. Simulations were conducted until the system reached steady state. Then, flame propagation speeds computed from the mass flow rate as well as flame shapes were extracted. The results are compared to those from \cite{kim_numerical_2006} for validation.
\begin{figure}[h!]
	\centering
	\includegraphics[width=9cm,keepaspectratio]{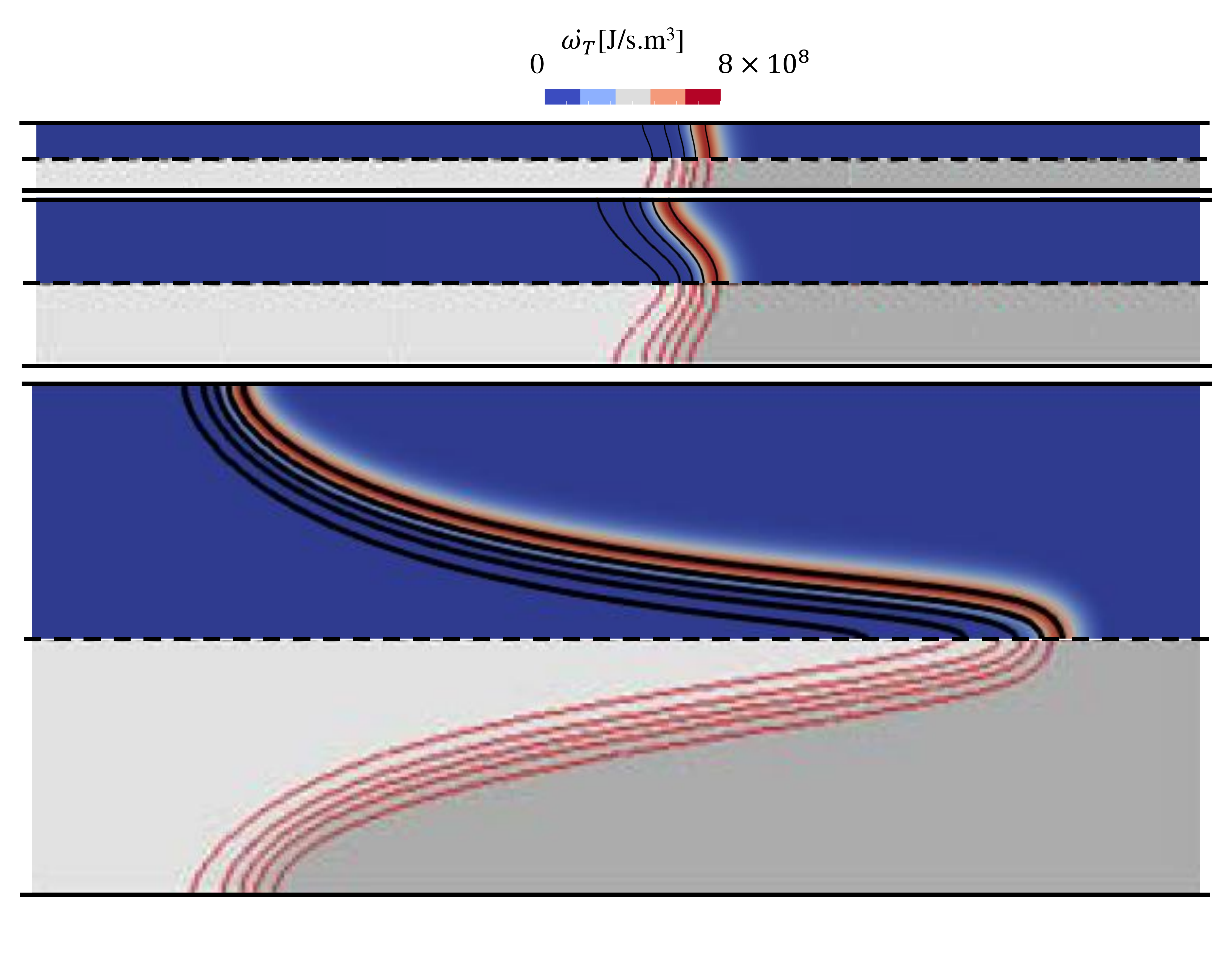}
	\caption{Comparison of flame shape obtained for adiabatic walls from simulations with ALBORZ (top half of each subfigure) to results from~\cite{kim_numerical_2006} (bottom half of each subfigure), with increasing channel height from top to bottom. The colors show the heat release rate, while the iso-contours (black in the top part, red in the bottom part) represent the following isotherms: $\theta=\frac{T-T_{\rm in}}{T_{\rm ad}-T_{\rm in}}\in\{0.1,\, 0.3,\, 0.5,\, 0.7,\, 0.9\}$. Reference images (bottom half of each subfigure) are reproduced from~\cite{kim_numerical_2006}. Channel widths are set to be true to scale.}
	\label{Fig:adiabatic_flame_shape}
\end{figure}
Starting from channels with widths comparable to the flame thickness (top part of Fig.~\ref{Fig:adiabatic_flame_shape}), deformations of the flame front due to the Poiseuille velocity profile are minimal. As the channel grows in width (from top to bottom in Fig.~\ref{Fig:adiabatic_flame_shape}) one observes more and more pronounced deformations at the center of the channel, effectively increasing the surface of the flame front. With more elongated flame surfaces one would expect changes in the propagation speed of the flame. The flame propagation speeds as a function of channel width are shown in Fig.~\ref{Fig:adiabatic_flame_speed} and again compared to reference data from~\cite{kim_numerical_2006}.
As a first observation it is seen that the present solver matches reference data very well. Furthermore, as expected from the changes in flame shape the flame propagation speed also increases with increased channel width, reaching speeds up to three time the laminar flame speed for $H=3$~mm.
\paragraph{Isothermal walls}
A second second set of simulations were then carried out while setting the wall boundary conditions to isothermal at $T_w = 300$~K. As for adiabatic walls, three different channel widths were considered, i.e. $H\in\{2.47,\, 3,\, 6\} {\rm mm}$. These channel widths were selected to cover the main flame shapes occurring for this configuration as expected from the literature, i.e. parabolic and tulip profile. The results obtained with ALBORZ are compared to simulations reported in~\cite{kim_numerical_2006} in Fig.~\ref{Fig:isothermal_flame_shape}.

\begin{figure}[h!]
	\centering
	\includegraphics[width=6cm,keepaspectratio]{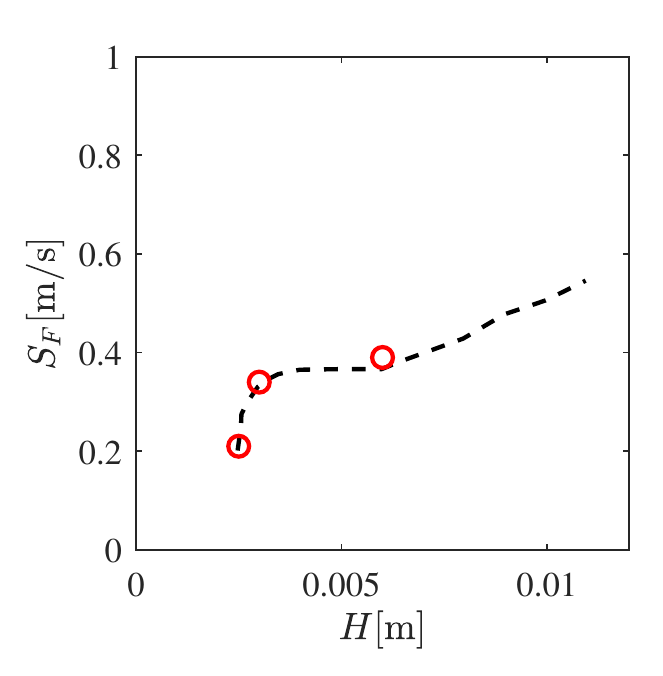}
	\caption{Comparison of flame propagation speed for isothermal walls obtained from  simulations with ALBORZ to results from~\cite{kim_numerical_2006} for different channel widths. Red circular markers are ALBORZ results while the black dashed line is data from ~\cite{kim_numerical_2006}.}
	\label{Fig:isothermal_flame_speed}
\end{figure}

\begin{figure}[h!]
	\centering
	\includegraphics[width=8cm,keepaspectratio]{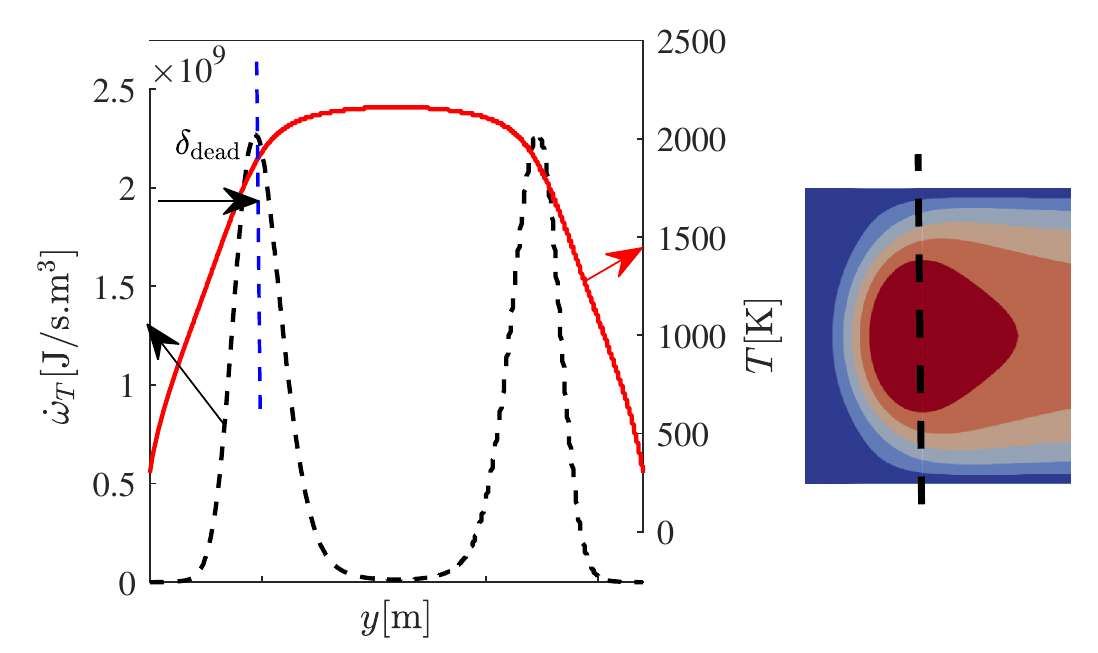}
	\caption{Illustration of dead zone minimum thickness for case with isothermal walls at $T_w=300$~K for $H=2.5$~mm. The right figure shows the flame in the channel (temperature field). The left figure corresponds to a cut through it along the dashed black line, plotting heat release (in black) and temperature (in red), together with the dead-zone limit (dashed blue line).}
	\label{Fig:deadspace_illustration}
\end{figure}

The results show good agreement with each other. Minor differences between results from ALBORZ and from~\cite{kim_numerical_2006} can, at least in part, be attributed to the fact that a two-step chemical mechanism is employed here, while~\cite{kim_numerical_2006} rely on a single-step, global mechanism. The propagation speeds were also extracted and compared to~\cite{kim_numerical_2006}, as shown in Fig.~\ref{Fig:isothermal_flame_speed}.
The agreement is observed to be very good for this quantity. Different from adiabatic walls where as channel width went down flame propagation speed converged to the free flame propagation speed, here as the channel width decreases the flame propagation speed goes below the free flame speed. This can be explained by the fact that lowering the channel width increases the energy loss toward the cold walls, compared to the energy released by the flame. It is also observed that at $H$ below $3~{\rm mm}$ the flame propagation speed  drops sharply; this corresponds to the onset of flame quenching discussed in the next paragraph.
\paragraph{Dead space and onset of quenching}
A closer look at Figs.~\ref{Fig:adiabatic_flame_shape} and \ref{Fig:isothermal_flame_shape} shows that the flame front hangs on to the walls for the adiabatic cases; On the other hand, for the isothermal cases there is a layer close to the walls where the flame is extinguished due to excessive heat losses, and fresh gas flow through; this zone is referred to as the dead zone~\cite{kim_numerical_2006}, as illustrated in Fig.~\ref{Fig:deadspace_illustration}.
Here, the quantity $\delta_{\rm dead}$ is introduced as the minimum thickness of the dead zone by monitoring the peak of heat release. To do that the position along the $x$-axis where the distance between the reaction front (marked by maximum of heat release) and the wall is minimum is found, and the corresponding distance along the normal to the wall is extracted. These values have been computed for four different cases (for the same widths as in the previous paragraph, and additionally for $H=2.1$~mm).
\begin{figure}[h!]
	\centering
	\includegraphics[width=5cm,keepaspectratio]{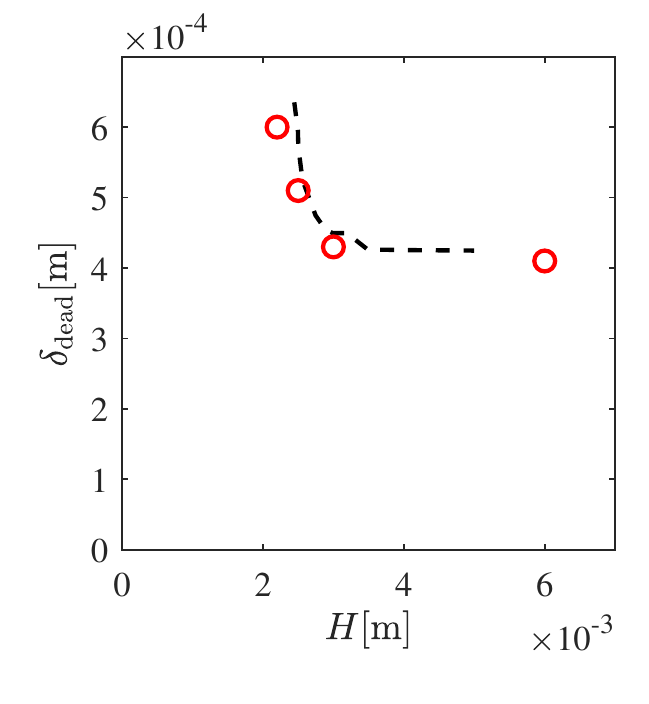}
	\caption{Comparison of dead zone minimum thickness for isothermal walls obtained from simulations with ALBORZ to results from~\cite{kim_numerical_2006} for different channel widths. Red circular markers are ALBORZ results while the black dashed line is data from~\cite{kim_numerical_2006}.}
	\label{Fig:deadspace_data}
\end{figure}
The results obtained with ALBORZ agree once more well with data from~\cite{kim_numerical_2006}. It is observed that for large channel widths the dead zone thickness reaches a lower plateau at a value of $\delta_{\rm dead}\approx 0.4~{\rm mm}$. As the channel width goes down the dead zone thickness experiences a rapid growth until the point where it becomes comparable to the channel width, so that the flame can not maintain itself anymore; this is called the quenching channel width. Calculations with ALBORZ led to a value of $H$ between 2 and $2.1~{\rm mm}$ for the quenching width, while~\cite{kim_numerical_2006} reported $H=2.4~{\rm mm}$. The difference between the two results can be probably attributed to the different chemical schemes employed, and grid- resolution, as reference uses an adaptive grid refinement procedure leading to grid sizes of $12.5~\mu{\rm m}$ in the diffusion and reaction layers; at such scales the slightest differences in laminar flame speed and thickness can have a pronounced effect on the flame/wall interaction dynamics.
\subsection{Methane/air premixed flame in pseudo 2-D reactor with cylindrical obstacles}
\begin{figure}[h!]
	\centering
	\includegraphics[width=6cm,keepaspectratio]{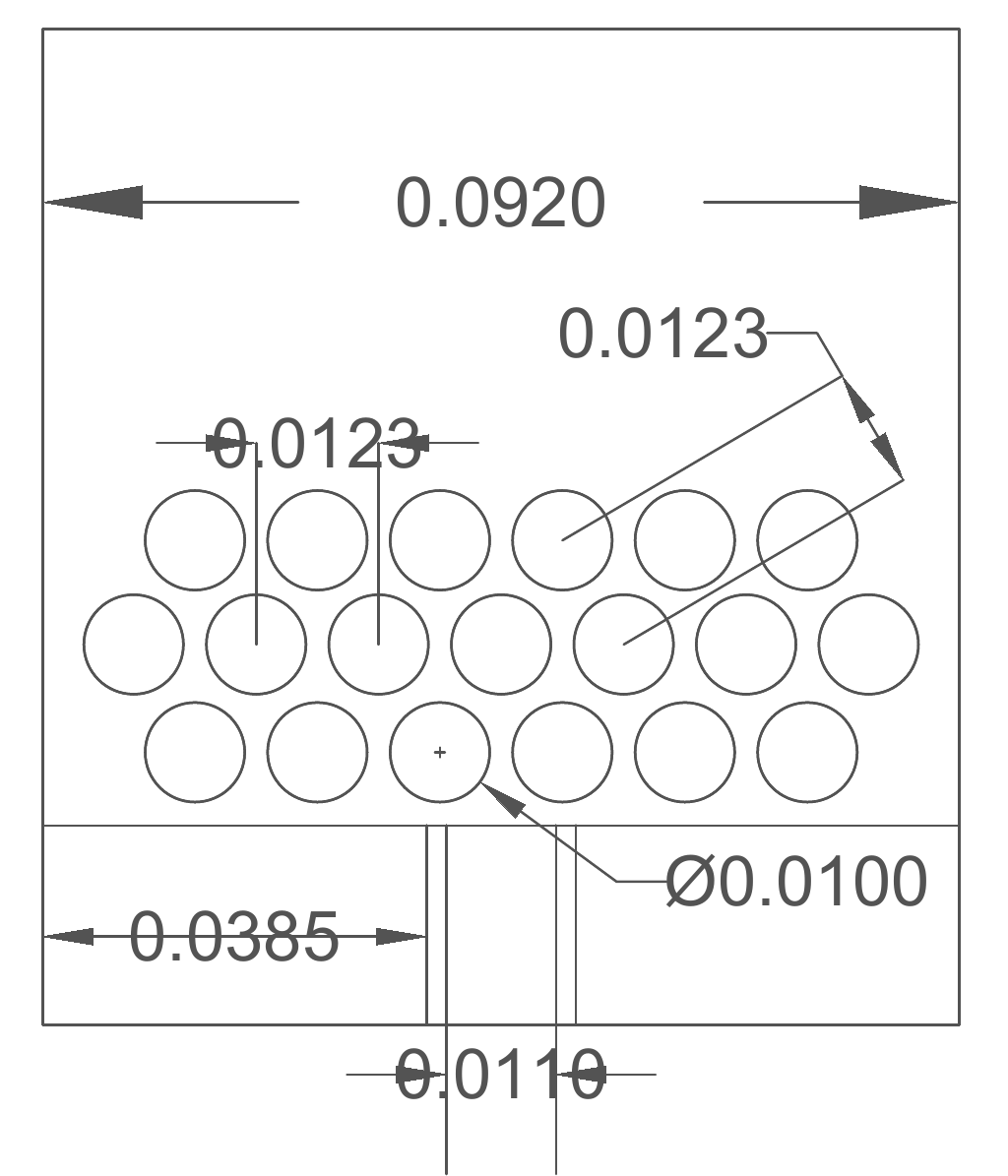}
	\caption{Geometry of the pseudo 2-D burner with cylindrical obstacles of~\cite{khodsiani_experimental_2021}.}
	\label{Fig:A3_2D_geometry}
\end{figure}

\begin{figure}[h!]
	\centering
	\includegraphics[width=6cm,keepaspectratio]{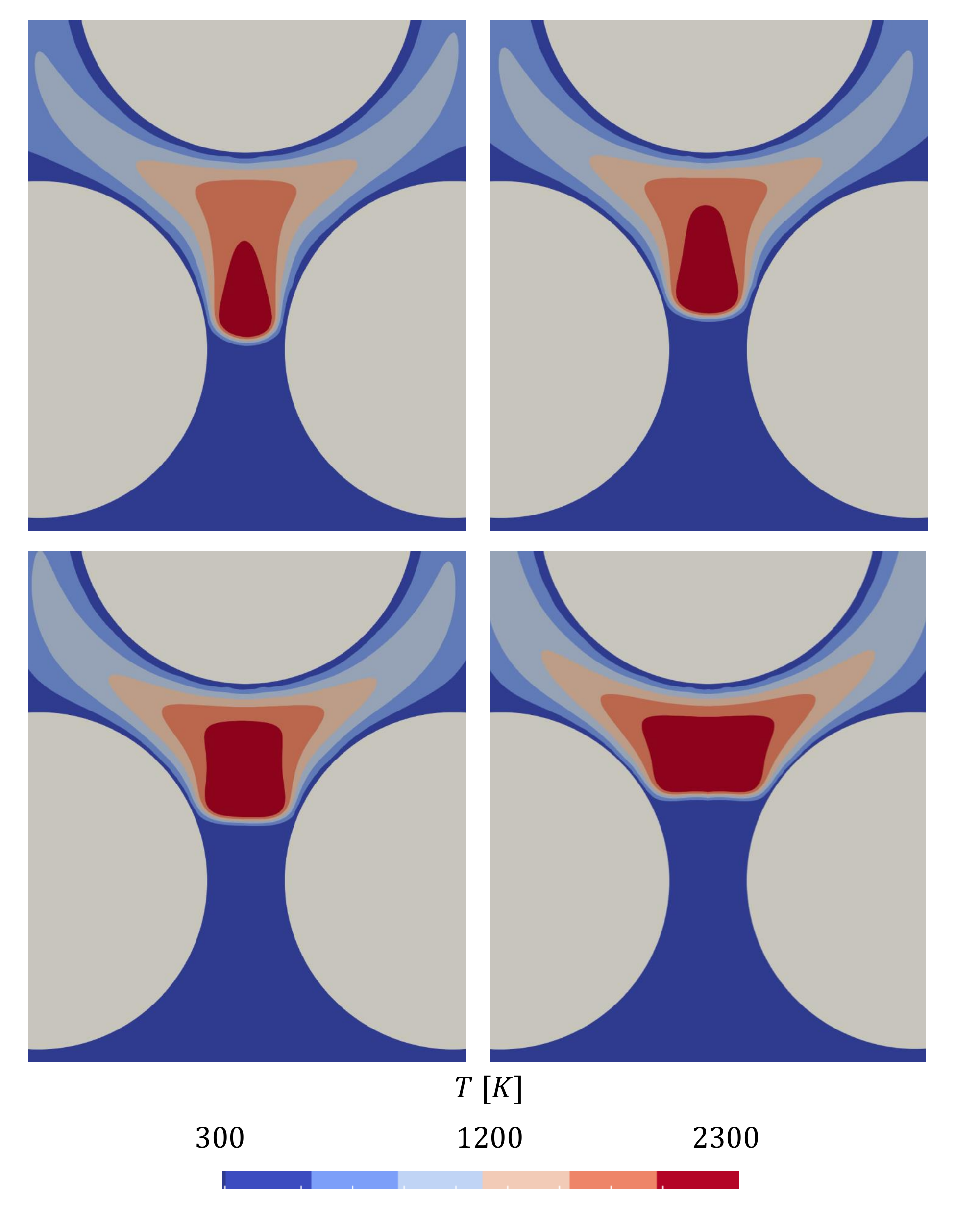}
	\caption{Evolution of the flame front as a function of time from top left to bottom right (corresponding to final steady state) in the configuration of Fig.~\ref{Fig:A3_2D_geometry}, illustrated via the temperature field.}
	\label{Fig:A3_flame_shape_evolution}
\end{figure}

\begin{figure}[h!]
	\centering
	\includegraphics[width=7cm,keepaspectratio]{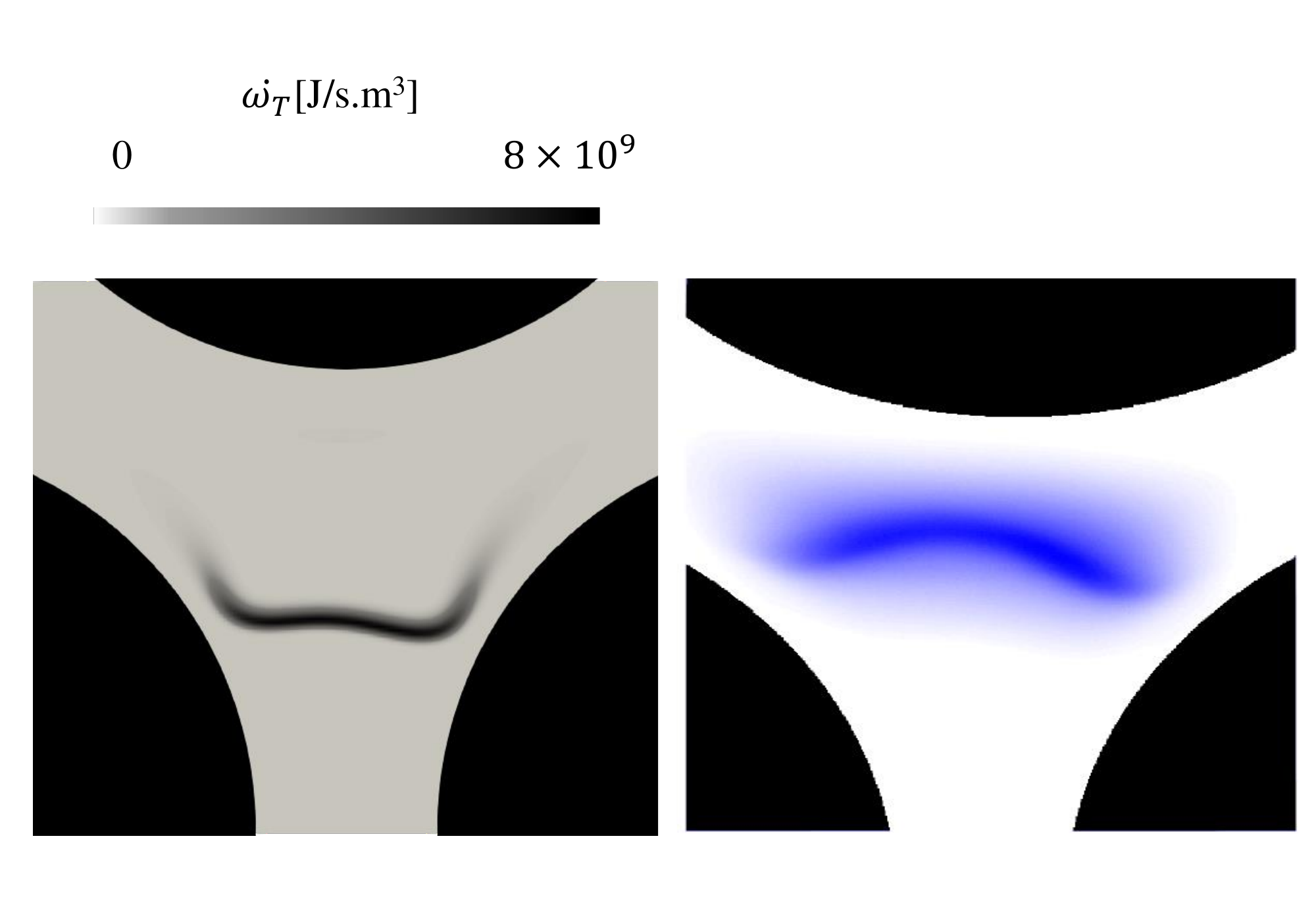}
	\caption{Illustration of flame front (right) reported in~\cite{khodsiani_flame_2023} from experiments compared to (left) simulations with ALBORZ for the geometry shown in Fig.~\ref{Fig:A3_2D_geometry}. The numerical image on the left shows the heat release rate, while the experimental image on the right captures all spontaneous emissions from species below 550~nm.}
	\label{Fig:A3_sym_flame_shape}
\end{figure}

The next case considered in this work is that of a pseudo-2D packed bed burner presented in \cite{khodsiani_experimental_2021}. It has been designed by colleagues at the University of Magdeburg in the Thermodynamics Group with the aim to replicate flow physics found in industrial packed beds by incorporating relevant size, geometry, and boundary conditions. For all details regarding design and measurement apparatus the interested readers are referred to~\cite{khodsiani_experimental_2021}. The overall geometry of the reactor, as initially intended, is illustrated in Fig.~\ref{Fig:A3_2D_geometry} in a vertical cut-plane through the center of the cylinders; it consists of a slit burner placed below a bed of cylindrical "particles". The rows of cylinders are arranged in an alternating pattern, with each consecutive row offset by precisely half the center-to-center distance. Most of the injected fuel/air mixture enters the packing between the two central cylinders of the first row, which are aligned with the slit burner.
The configuration considered involves a premixed methane/air mixture at equivalence ratio of one (stoichiometry) coming in from the central inlet at speed $0.3$~m/s, and air coming in from the two side inlets at the same speed to reduce the possible impact of external perturbations. All incoming fluxes are at temperature $25^\circ$C. All cylinders except three of them, the two central cylinders in the bottom-most row and the central cylinder in the middle row -- i.e., the three cylinders directly above fuel inlet, are associated to adiabatic no-slip walls as boundary conditions. The three remaining, central cylinders (the ones shown for instance in Fig.~\ref{Fig:A3_flame_shape_evolution}) are set to constant-temperature no-slip walls at $T_{\rm w}=373.15~{\rm K}=100^\circ$C, since they are thermostated at this particular temperature in the experiments. It should be noted that the measured temperatures in the experiment actually led to temperatures of $105\pm 1^\circ$C for the side cylinders and $120^\circ$C for the top central cylinder, which also might explain some of differences between simulation and experimental results. The simulations are conducted with resolutions $\delta r=0.05~{\rm mm}$ and $\delta t=0.1~\mu{\rm s}$.\\
Before looking at the steady position/shape of the flame and compare to experimental measurements, it is interesting to look at the unsteady evolution of the flame front and interpret these results based on the flame shapes discussed in the previous section. The flame evolution in the simulations is shown in Fig.~\ref{Fig:A3_flame_shape_evolution}.
The sequence of images present the flame front (described here by the temperature field) retracting along the positive $y$-direction going upward from the narrow gap between the two central cylinders in the bottom row toward the wider, inter-particle space located in-between the three isothermal cylinders facing the injection. In the narrowest cross-section (top-left image in Fig.~\ref{Fig:A3_flame_shape_evolution}) the flame shows a parabolic shape. As it moves further downstream, the center flattens and eventually goes toward a tulip shape (even better visible in Fig.~\ref{Fig:A3_sym_flame_shape}, left, showing heat release). Noting that at the narrowest section the equivalent channel width is $2.3$~mm, it can be seen that the behavior of the flame agrees qualitatively with that shown in Fig.~\ref{Fig:isothermal_flame_shape}(top) for the straight channel. At the widest section, i.e. for the bottom right snapshot in Fig.~\ref{Fig:A3_flame_shape_evolution}, $H\approx3.5$~mm. Referring again to the channel results discussed in the previous section, the flame front should be between a flattened parabola and a tulip (between middle and bottom row of Fig.~\ref{Fig:isothermal_flame_shape}), which is in good agreement with Figs.~\ref{Fig:A3_flame_shape_evolution} and~\ref{Fig:A3_sym_flame_shape} -- keeping in mind that the wall geometries are different in the channel and in the 2-D burner configurations.

\begin{figure}[h!]
	\centering
	\includegraphics[width=8.5cm,keepaspectratio]{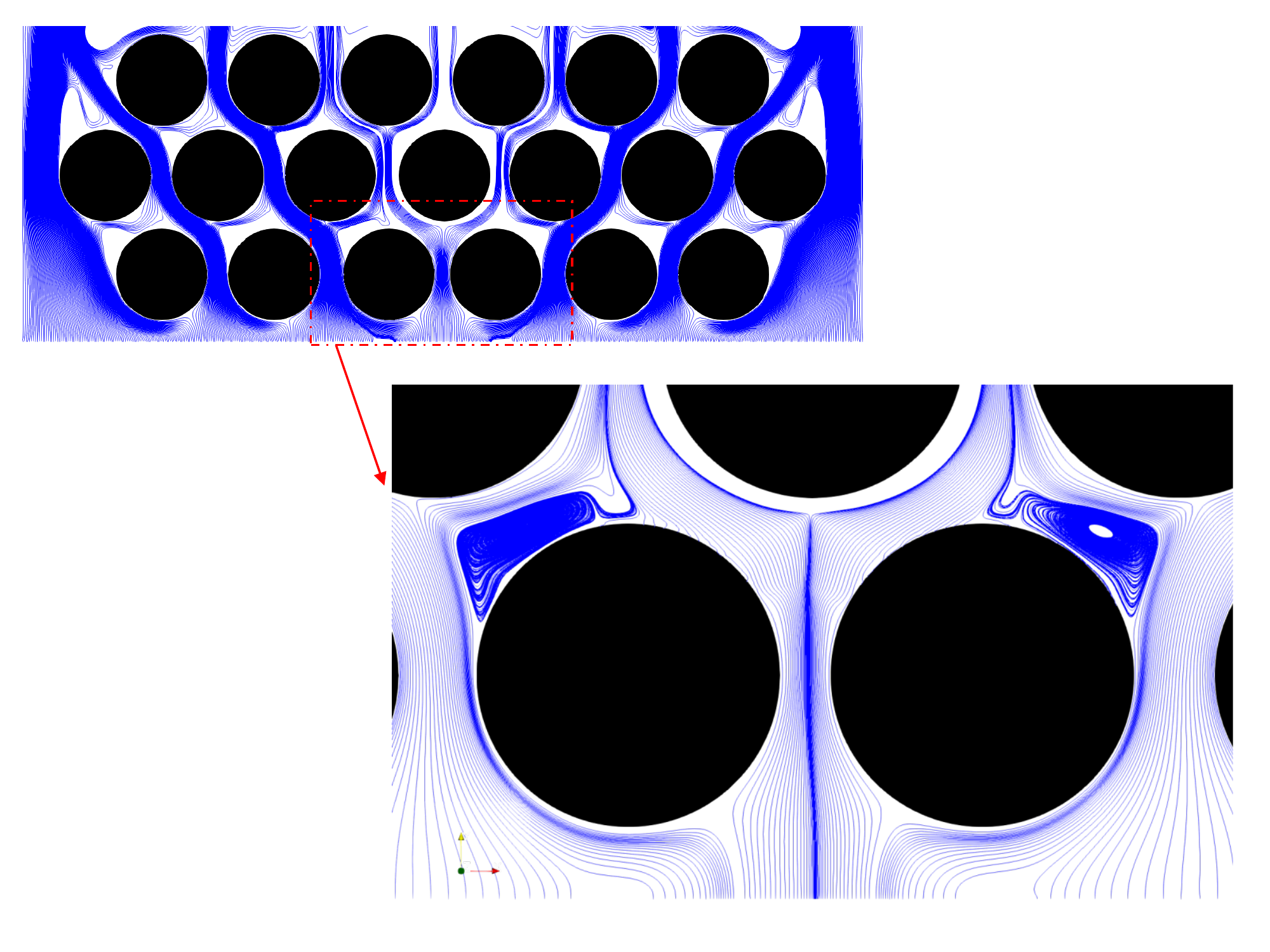}
	\caption{Flow structures illustrated by streamlines at steady state as obtained with ALBORZ based on the really measured geometry of the burner with cylindrical obstacles~\cite{khodsiani_flame_2023}.}
	\label{Fig:A3_asym_streamlines}
\end{figure}

\begin{figure}[h!]
	\centering
	\includegraphics[width=8.5cm,keepaspectratio]{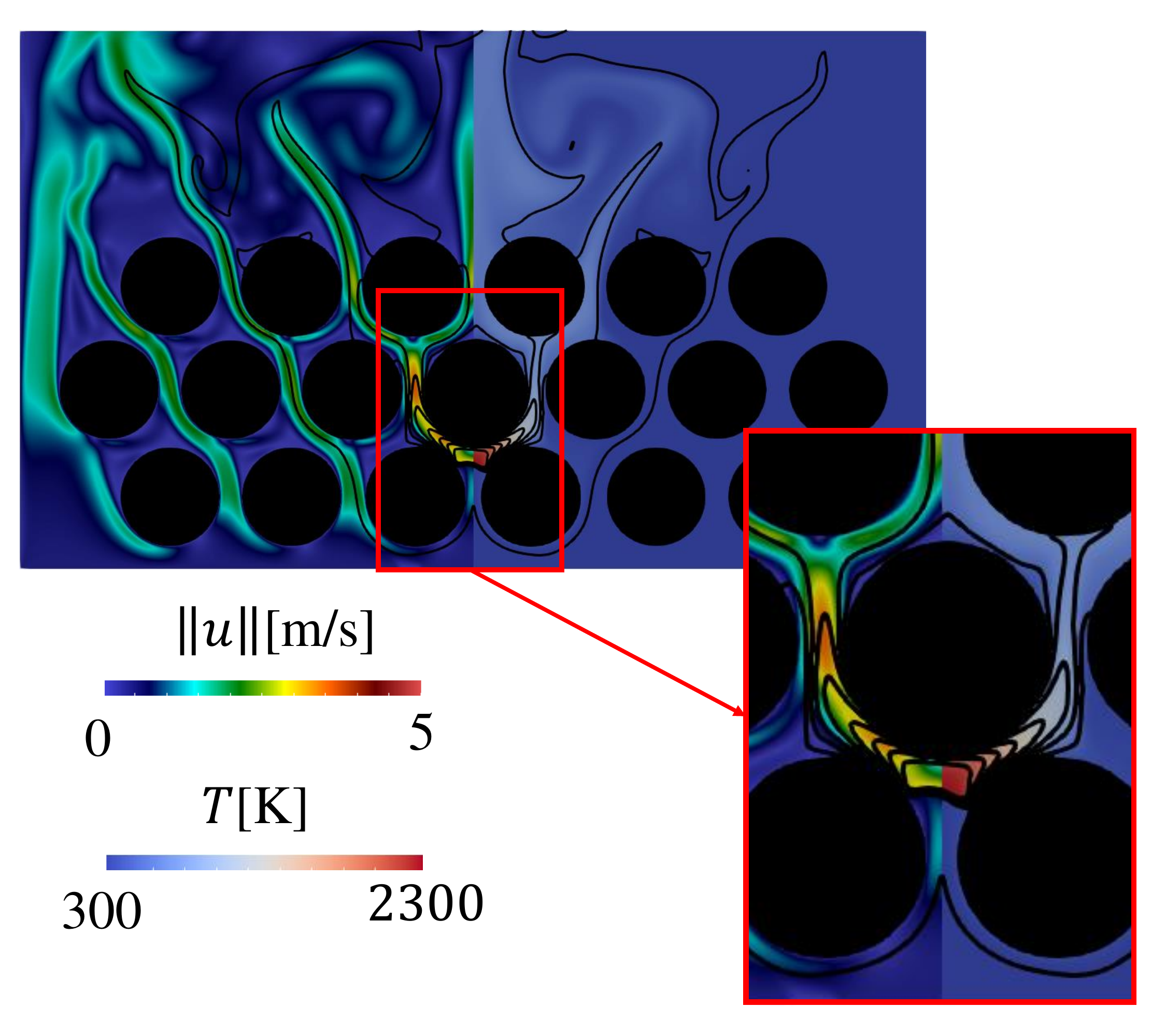}
	\caption{(Left half of each subfigure) Velocity magnitude and (right half of each subfigure) temperature fields in the full burner geometry obtained with ALBORZ based on the really measured geometry of the burner with cylindrical obstacles~\cite{khodsiani_flame_2023}. Iso-contours are for the temperature field dividing $T\in[300\,\,\,2300]$~K into 10 equally-spaced intervals.}
	\label{Fig:A3_asym_field}
\end{figure}

\begin{figure}[h!]
	\centering
	\includegraphics[width=7cm,keepaspectratio]{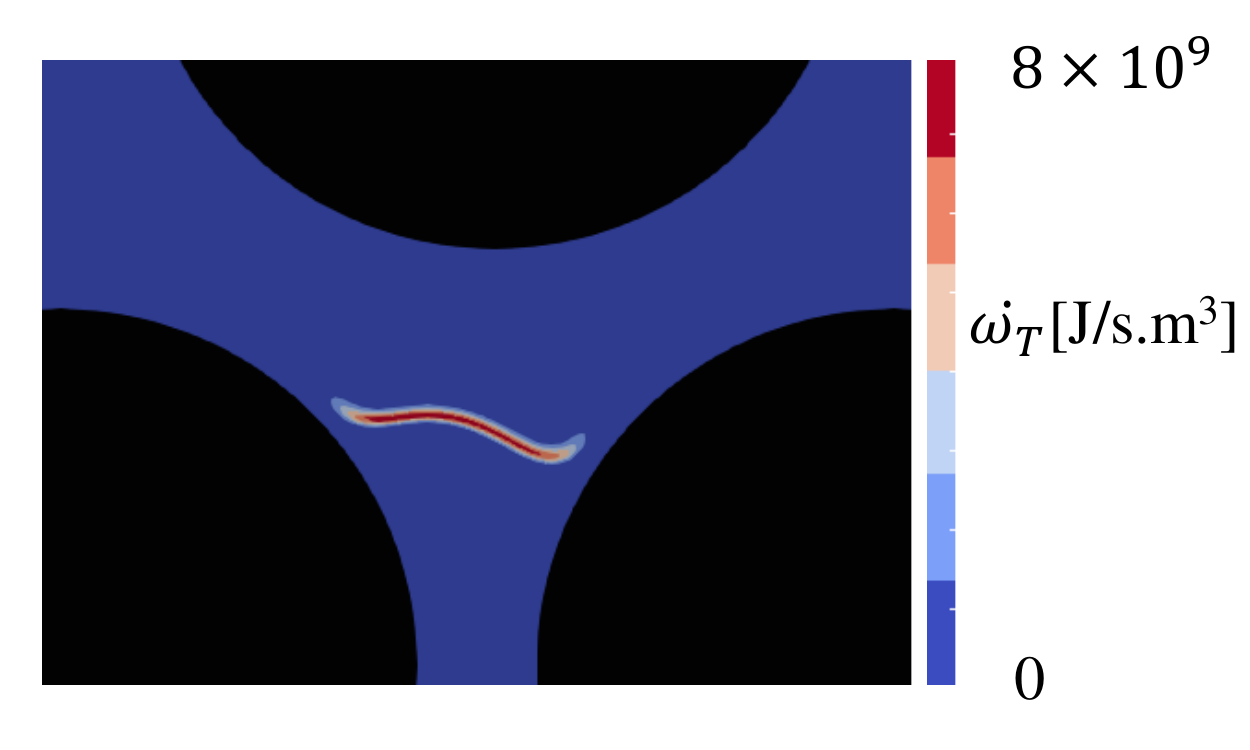}
	\caption{Flame shape and position illustrated via heat release as obtained from ALBORZ simulations for the really measured geometry.}
	\label{Fig:A3_asym_flame_shape}
\end{figure}

Furthermore, as for the channel with isothermal cold walls, the flame front exhibits a clear dead zone in regions neighboring the walls in Fig.~\ref{Fig:A3_flame_shape_evolution}, perhaps even better visible in  Fig.~\ref{Fig:A3_sym_flame_shape}(left). The flame front, as obtained from simulation, has been compared to experimental observations reported in~\cite{khodsiani_flame_2023} in Fig.~\ref{Fig:A3_sym_flame_shape}. In the experiments, the flame front is located at about 3.5~mm above the center of the first row of cylinders along the central vertical line, while in simulations it stabilizes at approximately 2.8~mm. Furthermore, experimental measurements point to an asymmetrical flame front.
This missing symmetry, as noted in~\cite{khodsiani_flame_2023}, might be possibly explained by small inaccuracies in the actual geometry of the burner compared to the design shown in Fig.~\ref{Fig:A3_2D_geometry}. To verify this point another simulation was carried out considering the finally \emph{measured} geometry of the real set-up as reported in~\cite{khodsiani_flame_2023}. The resulting flow field is illustrated via streamlines in Fig.~\ref{Fig:A3_asym_streamlines}. The streamlines at steady state show indeed a slightly asymmetrical flow configuration, especially in the region above the first row of cylindrical obstacles. Note that while the flow is unsteady above the bed, it reaches a steady configurations within.
The distribution of velocity and temperature in the full burner geometry is shown in Fig.~\ref{Fig:A3_asym_field}.
Th effect of the asymmetry in the flow field is better visible when looking at the flame front, shown in Fig.~\ref{Fig:A3_asym_flame_shape}.
Figure~\ref{Fig:A3_asym_flame_shape} shows that the asymmetrical flame shape observed in the experiments is better reproduced in the hybrid simulation when taking into account the really measured geometry. In particular, the flame becomes tilted, from top left to bottom right. Furthermore the flame stabilizes at a higher position, at 3.1~mm, matching better the experimental observations. The remaining discrepancy can be explained by different factors: minor differences in temperatures of iso-thermal cylinders as used in the simulation and as measured in experiments; non-homogeneous velocity and turbulence profiles at the inlet; and -- regarding simulations -- the simplicity of the chosen chemical scheme BFER-2, at the difference of a complete reaction mechanism. On top of this, while for simulations heat release was used to track the position of the flame front, experimental images contain spontaneous emissions from all species radiating below 550~nm, which is known to lead to a thicker flame front with deviations of the order of 0.1-1~mm regarding flame position toward the burnt gas region, i.e., here in streamwise direction, toward the top. Defining exactly the flame front has always been a challenge, since many different definitions are possible~\cite{Zistl}; this is even more true in experiments, considering that heat release can generally not be measured directly~\cite{Chi}. Keeping these points in mind, the agreement between experimental measurements and numerical results appears to be good. The obtained results already show a reasonable agreement between ALBORZ and measurement data, demonstrating that the numerical solver can well capture flow/flame/wall interactions. More detailed comparisons between experimental and numerical data will be the topic of future studies involving systematic parameter variations, and relying on additional quantities for the comparisons as soon as they have been measured experimentally. 
\subsection{Pore-resolved flame simulation in randomly generated porous media}
\begin{figure}[h!]
	\centering
	\includegraphics[width=8.5cm,keepaspectratio]{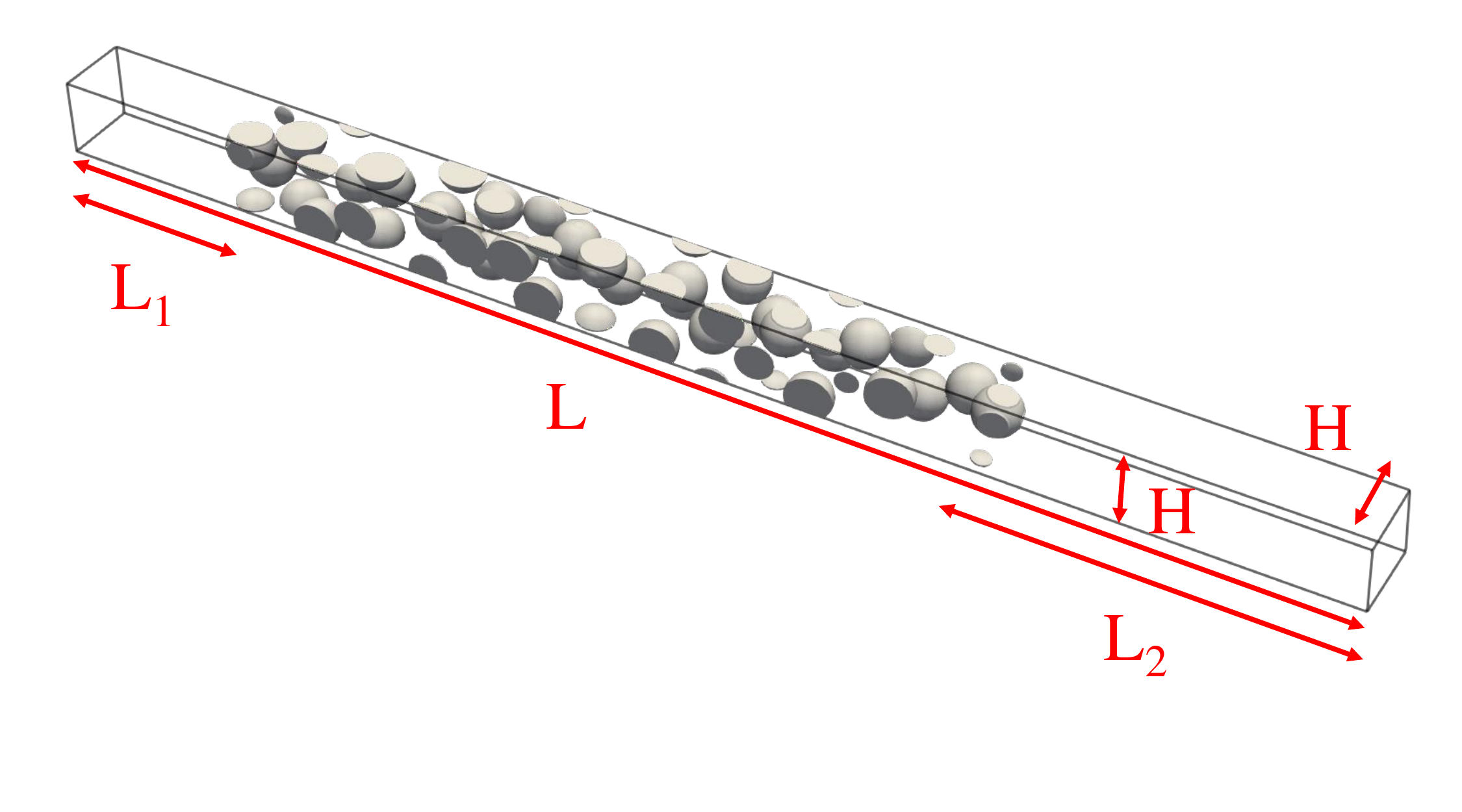}
	\caption{Illustration of randomly-generated porous media geometry.}
	\label{Fig:porous_geometry}
\end{figure}

\begin{figure}[h!]
	\centering
	\includegraphics[width=6cm,keepaspectratio]{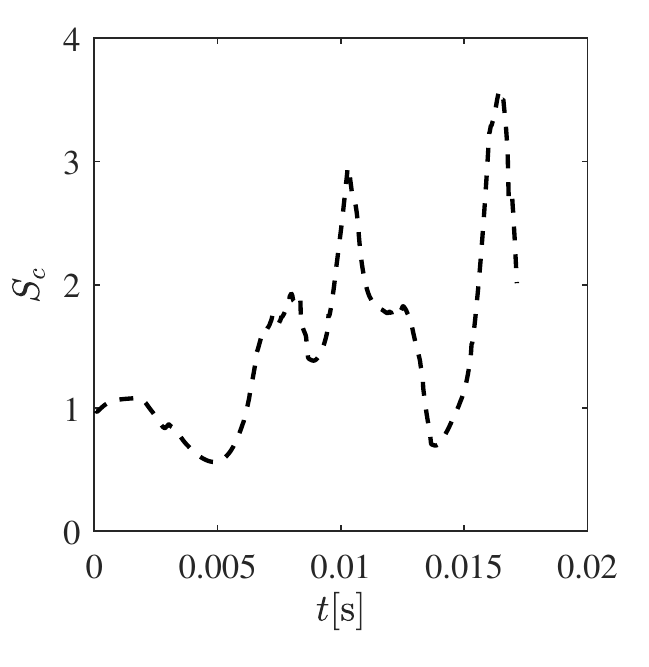}
	\caption{Evolution of methane consumption speed $S_c$ over time for flame propagation in porous media.}
	\label{Fig:porous_media_flame_speed}
\end{figure}

\begin{figure}[h!]
	\centering
	\includegraphics[width=8.5cm,keepaspectratio]{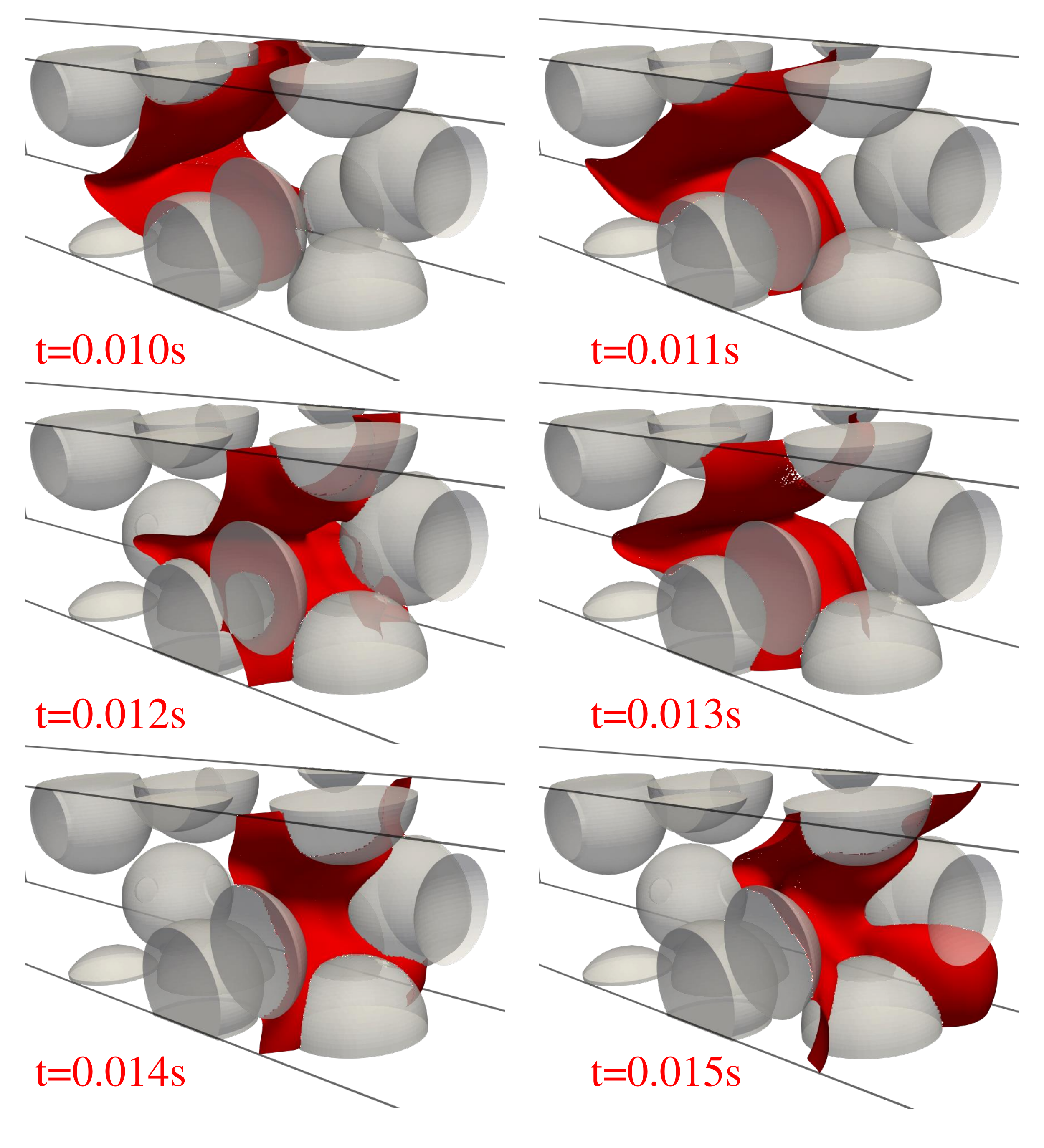}
	\caption{Illustration of the evolution of the flame at different times, represented by the temperature iso-surface $T=1500$~K.}
	\label{Fig:porous_flame_evolution}
\end{figure}

As a final configuration and to illustrate the applicability of the solver to more complex configurations, a geometry generated in the Porous Microstructure Analysis (PuMA) software~\cite{puma2018,puma2021} composed of randomly placed non-overlapping spheres with a diameter of 1.6 mm, a global porosity of 0.7 and a physical domain size $L\times H \times H$ with $L=0.08$~m  and $H=0.005$~m is considered. The geometry is illustrated in Fig.~\ref{Fig:porous_geometry}. Here $L_1 = 0.01$~m and $L_2=0.02$~m.
For this simulations the grid- and time-step sizes are set at the same values as in the previous configuration. Periodic boundary conditions are used for the top and bottom of the simulation domain. A constant mass flow rate boundary condition is used for the inflow (on the right), where the pressure and temperature are set to 1~atm and 298.15~K. At the inflow, the species mass fractions are set to that of the fresh gas at equivalence ratio 1. At the other end of the domain a constant hydrodynamic pressure along with zero-gradient boundary conditions for species and temperature field are used. During the simulation the total consumption speed of methane is monitored via:
\begin{equation}
    S_c = \frac{\int_V \dot{\omega}_{{\rm CH}_4} dV}{\int_V \dot{\omega}_{{\rm CH}_4}^{\rm flat} dV},
\end{equation}
where the consumption speed is normalized by that of a flat flame front, without any interaction with a porous media. The results are displayed in Fig.~\ref{Fig:porous_media_flame_speed}.
The average normalized propagation speed for this configuration is 1.797, with a large standard deviation of 0.6875. This larger propagation speed as compared to the laminar flame propagation speed is not unexpected. The flame dynamics in a porous media with adiabatic solid boundaries is mainly governed by the flame contortion as it goes over the solid obstacles. The consumption speed, in a process similar to that found for turbulent flames, is directly impacted by the increased flame surface. The evolution of the flame shape as it goes through the porous media is illustrated in Fig.~\ref{Fig:porous_flame_evolution}.
\section{Conclusions and discussion\label{sec:discussion}}
In this work a numerical model previously developed for gas-phase combustion has been extended and applied to reacting flows in porous media. Benchmark cases of increasing complexity in which flame/wall interactions dominate the dynamics of the system have been considered. It was shown that the model is able to capture the different flame/wall interaction regimes for both Dirichlet (constant temperature) and Neumann (adiabatic) boundary conditions. The suitability of the proposed solver for combustion simulations within a regular particle packing was discussed in connection to a pseudo 2-D burner involving cylindrical obstacles. First comparisons to experimental data point to a good agreement. Finally, for the first time to the authors' knowledge a lattice Boltzmann-based pore-scale simulation of combustion in a complex 3-D porous media is presented.
These results open the door for future studies considering flame propagation in realistic porous media and parametric studies of reacting gas flows in packed bed configurations.
\section*{Acknowledgement}
The authors acknowledge funding by the Deutsche Forschungsgemeinschaft (DFG, German Research Foundation) in TRR 287 (Project-ID 422037413), as well as the Gauss centre for providing computation time under grant "pn73ta" on the GCS supercomputer SuperMUC-NG at Leibniz Supercomputing Centre, Munich, Germany. Additionally, the authors thank Mohammadhassan Khodsiani, Beno\^it Fond and Frank Beyrau for interesting discussions regarding experimental measurements in the 2-D burner.
\section*{References}
\bibliography{references}
\end{document}